\documentclass[10pt,a4paper,onecolumn]{amsart}

\usepackage{mathtools}
\usepackage{amssymb}
\usepackage{amsfonts}
\usepackage{amsthm}
\usepackage{booktabs}

\usepackage{tikz}
\usepackage{pgfplots}

\usepackage{makecell}

\usepackage[foot]{amsaddr}
\makeatletter
\renewcommand{\email}[2][]{%
  \ifx\emails\@empty\relax\else{\g@addto@macro\emails{,\space}}\fi%
  \@ifnotempty{#1}{\g@addto@macro\emails{\textrm{(#1)}\space}}%
  \g@addto@macro\emails{#2}%
}
\makeatother

\usepackage{moreverb}
\usepackage{cite}

\usepackage{todonotes}

\newcommand{\face}{\ensuremath{{\!f}}}
\newcommand{\own}{\ensuremath{{\!P}}}
\newcommand{\nei}{\ensuremath{{\!N}}}
\newcommand{\cen}{\ensuremath{{\!C}}}
\newcommand{\dow}{\ensuremath{{\!D}}}
\newcommand{\upw}{\ensuremath{{\!U}}}
\newcommand{\grad}{\ensuremath{{\nabla\!}}}

\usepackage[dvips,colorlinks,bookmarksopen,bookmarksnumbered,citecolor=red,urlcolor=red]{hyperref}
\usepackage{subcaption}
\usepackage[percent]{overpic}

\newcommand\BibTeX{{\rmfamily B\kern-.05em \textsc{i\kern-.025em b}\kern-.08em
T\kern-.1667em\lower.7ex\hbox{E}\kern-.125emX}}
\newcommand{\scheme}[1]{#1}

\begin{document}

\title[Implicit mesh-skewness correction for VOF based mass transfer]{Boundedness-preserving implicit correction of mesh-induced errors for VOF based heat and mass transfer}

\author{S.~Hill\textsuperscript{1,3}}
\author{D.~Deising\textsuperscript{2}}
\author{T.~Acher\textsuperscript{1}}
\author{H.~Klein\textsuperscript{3}}
\author{D.~Bothe\textsuperscript{2}}
\author{H.~Marschall\textsuperscript{2,*}}
\email{Corresponding Author: \textsuperscript{*}\href{mailto:marschall@mma.tu-darmstadt.de}{marschall@mma.tu-darmstadt.de}}

\address[1]{Linde Engineering AG, Pullach, Germany}
\address[2]{Mathematical Modeling and Analysis, Fachbereich Mathematik, Technische Universit\"at Darmstadt, Darmstadt, Germany}
\address[3]{Institute of Plant and Process Technology, Faculty of Mechanical Engineering, Technical University of Munich, Munich, Germany}

\begin{abstract}
Spatial discretisation of geometrically complex computational domains often entails unstructured meshes of general topology for 
Computational Fluid Dynamics (CFD).
Mesh skewness is then typically encountered causing severe deterioration of the formal order of accuracy of the discretisation, 
or boundedness of the solution, or both. Particularly methods inherently relying on the accurate and bounded transport of sharp 
fields suffer from all types of mesh-induced skewness errors, namely both non-orthogonality and non-conjunctionality errors.

This work is devoted to a boundedness-preserving strategy to correct for skewness errors arising from discretisation of advection and diffusion 
terms within the context of interfacial heat and mass transfer based on the Volume-of-Fluid methodology. The implementation has been accomplished 
using a second-order finite volume method with support for unstructured meshes of general topology.
We examine and advance suitable corrections for the finite volume discretisation of a consistent single-field model, where both accurate and bounded transport 
due to diffusion and advection is crucial. In order to ensure consistency of both the volume fraction and the species concentration transport, 
i.e.\ to avoid artificial heat or species transfer, corrections are studied for both cases. The cross interfacial jump and adjacent sharp 
gradients of species concentration render the correction for skewness-induced diffusion and advection errors additionally demanding and 
has not so far been addressed in the literature.

\end{abstract}


\maketitle

\vspace{-6pt}

\section{Introduction}
\vspace{-2pt}

Algebraic Volume-of-Fluid (VoF) methods \cite{Ubbink1997, Muzaferija1998, Ubbink1999, Muzaferija1999} are widely used for numerical simulations of flows consisting of two incompressible fluid phases with a sharp deformable interface separating them. Recently a model for interfacial species transfer within the algebraic VoF framework, namely the Continuous Species Transfer (CST) model, 
was presented in \cite{Marschall2012,Deising2016}. Within this modeling framework, heat transfer is to be seen as a special case without jump and can thus be described by a simplified CST model equation. Both methods, VoF and CST, are derived by applying the Conditional Volume-Averaging (CVA) technique to the local instantaneous sharp interface model equations and the corresponding interfacial jump conditions so as to transform them into volume-averaged single-field equations. In this form the governing equations are readily suitable for consistent discretisation using a Finite Volume Method (FVM). 

\begin{figure}[htb]
  \centering
    \begin{subfigure}[b]{0.49\textwidth}
    \centering
      \includegraphics[width=0.9\textwidth]{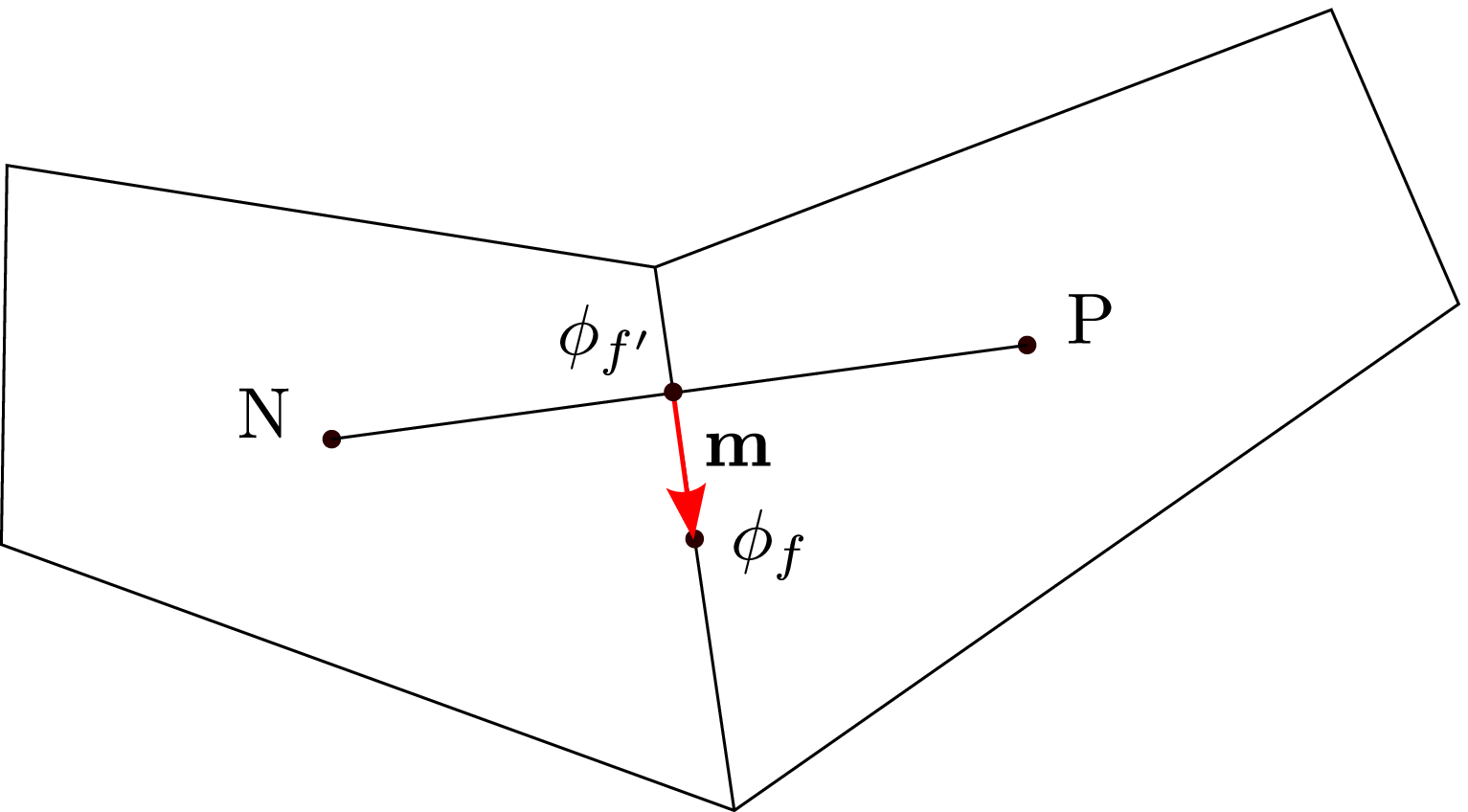}
    \caption{Non-conjunctionality.}
    \label{figNCMesh}
    \end{subfigure}
  \begin{subfigure}[b]{0.49\textwidth}
    \centering
    \includegraphics[width=0.9\textwidth]{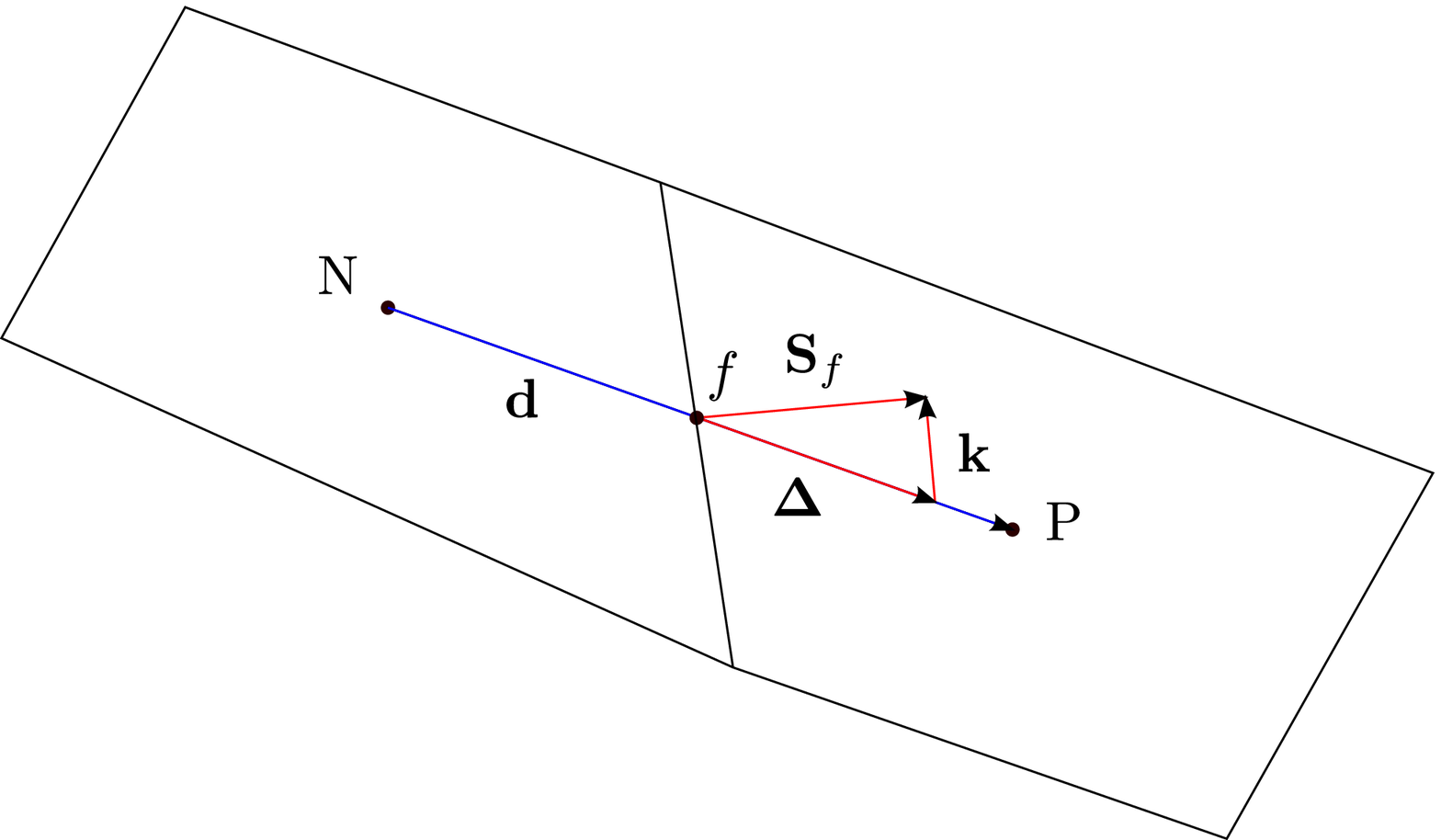}
    \caption{Non-orthogonality.}
    \label{figNOMesh}
  \end{subfigure}
  \caption{Types of mesh skewness.}
  \label{figSkewedMesh}
\end{figure}

When considering computational domains entailing complex geometries, unstructured meshes of general topology are required. Then, depending on the mesh quality, discretisation errors related to mesh skewness arise and can severely detoriate the numerical fidelity of the method resulting in loss of boundedness, accuracy and order of convergence. Typically, two types of skewness errors are to be distinguished: errors due to non-conjunctionality and errors due to non-orthogonality at the three-point stencil around a cell face (see Figure \ref{figSkewedMesh}). In order to avoid detoriation of numerical fidelity on such meshes, skewness correction approaches have been devised for FV discretisation and are nowadays well established practice. Classical approaches are explicit and correct for non-orthogonality during discretisation of diffusion terms and non-conjunctionaly during FV discretisation of advection terms, respectively. For an overview of state-of-the-art classical approaches to skewness correction the interested reader is referred to \cite{Jasak1996, Ferziger2002, Moukalled2015}. A non-classical explicit approach is the ghost-point method set out by Peri{\'c} in \cite{Ferziger2002}, correcting for both non-conjunctionality and non-orthogonality simultaneously.

Applying classical (explicit) skewness correction approaches to the VoF advection equation on distorted meshes results directly in severe unboundedness of the volume fraction and as a consequence in substantial problems with stability and/or accuracy. The reason for this resides in the fact that due to the typically high density ratios even small errors in the volume fraction cause large errors in mass fraction and consequently in the momentum. It has been found that in such cases implicit correction approaches are needed, see \cite{Croft1998, Zhang2006, Zhang2007} and more recently \cite{Denner2014, Denner2015}. Taking interfacial heat and mass transfer into consideration renders the situation even more challenging, since there are now both the discretised advection and diffusion terms to be skewness corrected. While it is obvious that classical explicit approaches remain unsuitable (likewise causing unboundedness errors as for the VoF transport), up to now it is unclear whether implicit correction approaches can be used to correct for mesh skewness when applied to the discretised form of the CST model for interfacial heat and mass transfer. 

This contribution is concerned with enhancements of implicit correction approaches based on a detailed investigation into their suitability to maintain numerical fidelity such as boundedness, accuracy and order of convergence in the context of VoF/CST-based simulations of interfacial heat and mass transfer on distorted meshes of general topology. In contrast to previous work, we introduce implicit skewness corrections for finite volume discretisations of both advection and diffusion terms. Moreover, we use finite volume meshes which have been distorted in a targeted manner rather than randomly (as is common practice in the existing literature) in order to allow a systematic study on mesh-induced errors and avoid error cancellation. The objectives of this work are two-fold, viz.\
\begin{itemize}
    \item to devise a suitable method for mesh skewness correction for the discretised advection and diffusion terms of the single-field CST model for VoF-based 
        interfacial heat and mass transfer. Notably, one of the CST model terms is discretised in a non-standard way \cite{Deising2016}, that is, despite originating 
        from a diffusive transport term, it has been found that it is best to treat one term implicitly by means of a FV divergence rather than a FV laplacian operator.
    \item to identify and clarify the influence of a jump in the scalar transport variable on the order of convergence when refining a systematically distorted 
        finite volume mesh and applying implicit skewness corrections.
\end{itemize}
This study has been accomplished using and extending OpenFOAM (version 4.x).

\section{Standard finite volume discretisation and Mesh-Induced Errors}
\label{sec:finiteVolume}
\vspace{-2pt}
In order to better understand the proposed approach to implicit skewness correction for VoF/CST-based interfacial heat and mass transfer on distorted meshes of general topology,
the errors on skewed meshes in finite volume discretisation of a prototypical scalar advection-diffusion equation are briefly recapped and a review of prominent publications addressing these errors is given. 

The starting point of our investigations is the advection-diffusion equation for an arbitrary scalar quantity $\phi$, i.e.\
\begin{equation}
    \partial_t\phi+\nabla\cdot\left(\phi \mathbf{u}\right)
   =-\nabla\cdot\left(-\Gamma_{\!\phi}\grad\phi\right) + S_{\!\phi} \, .
    \label{eqnLocal}
\end{equation}
In the finite volume context, the standard discretisation procedure is based on the application of Gauss' rule,
transferring the divergence terms into face integrals, and on the mid-point rule. This yields a formal second-order accurate approximation. 
In collocated finite volume methods, as employed in this work, the values of quantity $\phi$ are stored in the cell centre locations,
which necessitates the interpolation of cell-centred data onto the face centres. Below cell-face interpolation will be denoted by $\left(\cdot\right)_{\face}$.
In order to keep the discretisation second order accurate, the interpolation schemes have to be at least of the same order.  
However, in general, even if higher order interpolation schemes are selected, the convergence order as well as the overall accuracy can still deteriorate, particularly in cases of complex computational domain geometries and meshes of general topology (polyhedral meshes).  The errors stem directly from local mesh skewness, namely non-orthogonality and non-conjunctionality. They can be corrected during discretisation of the respective transport terms in \eqref{eqnLocal}.

\subsection{Advection term}
The standard finite volume discretisation of advective terms reads
\begin{equation}
\begin{aligned}
    \int_V\mathbf \nabla\cdot\left(\phi \mathbf{u}\right)\,dV 
    &= \oint_S\mathbf n\cdot\left(\phi \mathbf{u}\right)\,dS
    \approx \sum\limits_\face \mathbf S_{\face}\cdot\left(\phi \mathbf{u}\right)_{\face} \\
    &\approx \sum\limits_\face \mathbf S_{\face}\cdot\mathbf{u}_{\face} \phi_{\face}
    =: \sum\limits_\face F_{\face} \phi_{\face}  \, ,
    \label{eqnAdvect}
\end{aligned}
\end{equation}
with the volumetric face flux $F_{\face}:=\mathbf S_{\face}\cdot\mathbf{u}_{\face}$. The face-centred values of $\phi$ are obtained by using interpolation schemes. In general, second order accuracy of the discretisation formally relies on the assumption that the face-centre position corresponds to the intersection point of the face and the connecting line between the neighbouring cell centres. If this assumption does not hold, as the constellation depicted in Figure~\ref{figNCMesh} shows by way of example, 
so called \textit{non-conjunctionality errors} occur. 

The standard approach to account for this error is to correct the face-interpolated value $\phi_{\face'} = f(\phi_{\nei},\phi_{\own})$ by an explicit term reading
\begin{equation}
    \phi_{\face}=\phi_{\face'}+\nabla\!\phi_{\face'}\cdot\mathbf m \, ,
    \label{eqnSkew}
\end{equation}
where $\mathbf m$ denotes the vector from the face intersection point to the face centre.

The above time-explicit correction of the mesh-induced non-conjunctionality error
is very simple and computationally inexpensive but may lead to unboundedness of the transported quantity \cite{Jasak1996},
e.g. when high-resolution schemes (based on TVD, NVD face limiters etc.) are utilised for the discretisation
as is common practice in FVM.
Magalhaes et al.~\cite{Magalhaes2013} introduce a similar approach for adaptive hexahedral meshes
to account for non-conjunctionality errors, where the face value is reconstructed from the face-neighbouring 
cell-centred values and the respective gradients, and then averaged:
\begin{equation}
    \phi_{\face} = \frac{1}{2} \left( \left[ \phi_{\own} + \grad\phi_{\own} \cdot \left( \mathbf{x}_\face - \mathbf{x}_\own \right) \right]
           + \left[ \phi_{\nei} + \grad\phi_{\nei} \cdot \left( \mathbf{x}_\face - \mathbf{x}_\nei \right) \right] \right) \, .
    \label{eqnSkewAlternative1}
\end{equation}
An entirely different correction approach, the so-called \emph{ghost point-based interpolation scheme},
which corrects the face interpolation by introducing a virtual non-skew stencil, was proposed by Ferziger and Peric \cite{Ferziger2002} and successfully employed e.g.\ in 
\cite{Albuquerque2013}. However, if boundedness of the transported quantity is of utmost importance, as e.g.\ in the VoF-based transport of the volumetric phase fraction or species concentration fields, the above \emph{explicit} methods are not usable as they do not ensure this property.

Therefore, this work makes use of a modified \emph{implicit} version of the standard correction approach described by Equation~({\ref{eqnSkew}), 
which makes use of an improved method for preserving the boundedness of the transported quantity and is described in detail in Section \ref{sec:BoundednessCriteria}.

\subsection{Diffusive term}
Analogously to Equation (\ref{eqnAdvect}), the discretised form of the diffusive term is given by
\begin{equation}
    \oint_S\mathbf n\cdot\left(\Gamma_{\!\phi}\grad\phi\right)\,dS
    \approx\sum\limits_\face \mathbf S_\face\cdot\left(\Gamma_{\!\phi}\grad\phi\right)_\face
    \approx\sum\limits_\face \left(\Gamma_{\!\phi}\right)_\face\mathbf S_\face\cdot\left(\grad\phi\right)_\face\,.
    \label{eqnDiffus}
\end{equation}
The term $\left(\Gamma_{\!\phi}\right)_\face\mathbf S_\face\cdot\left(\grad\phi\right)_\face$ is preferably to be discretised in a compact way, that is only information of the
two cells neighbouring the respective face is considered:
\begin{equation}
  \left(\Gamma_{\!\phi}\right)_\face\mathbf S_\face\cdot\left(\nabla\phi\right)_\face
  =\left(\Gamma_{\!\phi}\right)_\face|\mathbf S_\face|\nabla_\face^\perp\phi
	=\left(\Gamma_{\!\phi}\right)_\face|\mathbf S_\face|\frac{\phi_\nei-\phi_\own}{|\mathbf d|} \, .
  \label{eqnSnGrad}
\end{equation}
This approximation of the \textit{surface normal gradient} $\nabla_\face^\perp\phi$, however, only holds true for orthogonal meshes.
As a consequence, a numerical error is introduced in case of local non-orthogonality, i.e.\ if $\mathbf S_\face\nparallel \mathbf d$ 
(see Figure~\ref{figNOMesh}). 

One way to account for this error was introduced in \cite{Jiang1994} and \cite{Davidson1996}, where cell vertex-interpolated values,
obtained by either inverse distance weighting or least-squares interpolation, are utilized
to correct for mesh non-orthogonality, which in effect significantly increases the numerical stencil.
This, however, counteracts the accurate discretization of discontinuities, which are inherently smoothed
by a large discretization stencil.

A different approach, proposed by \cite{Demirdzic1990,Demirdzic1995,Mathur1997}, is to split up the vector $\mathbf S_\face$ in such a way that 
\begin{equation}
    \mathbf S_\face=\mathbf\Delta + \mathbf k\,,
    \label{eqnSf}
\end{equation}
with the restriction of $\mathbf\Delta$ being  parallel to $\mathbf d$. Employing the above splitting, the discretisation can be performed as
\begin{equation}
    \mathbf S_\face\cdot\left(\grad\phi\right)_\face
    =\mathbf \Delta\cdot\left(\grad\phi\right)_\face + \mathbf k\cdot\left(\grad\phi\right)_\face
    =|\mathbf\Delta|\frac{\phi_\nei-\phi_\own}{|\mathbf d|} + \mathbf k\cdot\left(\grad\phi\right)_\face \,.
    \label{eqnSnGradNO}
\end{equation}
For the computation of the magnitude of $\mathbf\Delta$ and $\mathbf k$ different options are available 
(Jasak~\cite{Jasak1996}, Demirdzic~\cite{Demirdzic2015}).
For further reading regarding non-orthogonal corrections the interested reader is also referred to 
the works of Demirdzic~\cite{Demirdzic1990,Demirdzic1995,Demirdzic2015}, as well as Mathur and Murthy~\cite{Mathur1997}.
These pionieering works represent the main contributions to non-orthogonality corrections and their development history.
An overview of alternative but similar correction approaches can also be found in \cite{Segarra2004}.
All of the above mentioned approaches eventually result in a formulation similar to or exactly the same as that given in \eqref{eqnSnGradNO}.
\begin{figure}[htb]
   \centering
   \includegraphics[width=0.45\textwidth]{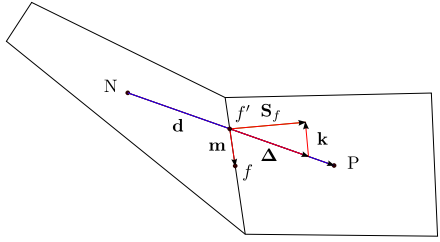}
   \caption{Generally skewed unstructured mesh due to non-conjunctionality and non-orthogonality.}
   \label{figSkewedNOMesh}
\end{figure}
If in addition the mesh exhibits local non-conjunctionality, i.e.\ is generally skewed (see Figure~\ref{figSkewedNOMesh}), further corrections may have to be applied.
One available explicit method for correcting both skewness errors simultaneously is the ghost-point interpolation proposed by \cite{Ferziger2002}.
Generally, non-conjunctionality corrections in the discretisation of diffusion terms are simply omitted without comment. In this work, however, their influence on accuracy will be investigated.

\vspace{-6pt}

\section{Proposed Bounded Implicit Correction of Mesh-Induced Errors}
\label{sec:BoundednessCriteria}
\vspace{-2pt}
As the explicit correction of mesh-induced errors can lead to unboundedness, suitable implicit corrections must be devised. For this purpose, we first briefly recap boundedness-preserving criteria from the literature and then adapt the formulation of the skewness correction in a way such that they can be applied. 

\subsection{Boundedness Criteria of Advection Term on Unstructured Meshes}
\label{ssec:BoundednessCriteriaAdvection}
To enforce boundedness during discretisation of the advection term, different boundedness criteria based on \textit{positivity} \cite{Hirsch2007}, are readily available in literature. Most prominent (inter alia) are the
\textit{Total Variation Diminishing (TVD)} \cite{Harten1983} and \textit{Local Extrema Diminishing (LED)} \cite{Jameson1995} criteria
as well as the \textit{transient Convection Boundedness Criterion (CBC)} \cite{Leonard1988}.
\begin{figure}[htb]
    \centering
    \includegraphics[width=0.6\textwidth]{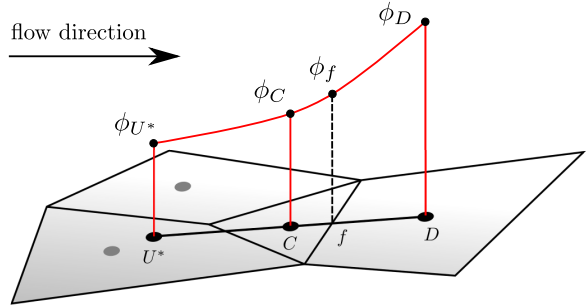}
    \caption{Considered stencil for discretisation.}
    \label{figStencil}
\end{figure}

Defining an interpolation scheme based on a flux limiter formulation as \cite{Sweby1984,Harten1983}
\begin{equation}
    \phi_\face=\phi_\cen+\delta_\face \Psi\!\left( r \right)\, \left(\phi_\dow-\phi_\cen\right),
    \label{eqnInterpolationScheme}
\end{equation}
with the mesh weight $\delta_\face = \frac{\mathbf{d}_{\!\text{Cf}}\cdot\mathbf{S}_\face}{\mathbf{d}_{\!\text{CD}}\cdot\mathbf{S}_\face}$ and the 
limiter function $\Psi\!\left(r\right)$, the TVD conditions can be transferred into 
\begin{equation}
    0 \leq \Psi\!\left(r\right) \leq \min\left(\frac{1}{\delta_\face}r\frac{1-\operatorname{Co}_\face}{\operatorname{Co}_\face}, \frac{1}{\delta_\face}\right) \, ,
    \label{eq:TVDconditionTransferred}
\end{equation}
with $\operatorname{Co}_\face$ denoting the face Courant number and $r$ the gradient ratio, being defined as 
\begin{equation}
    \operatorname{Co}_\face=\left( \frac{\Delta t}{|V|} \sum\limits_\face \max \left( F_\face, 0 \right) \right)_{\face,\text{UD}}  \quad , \quad r=\frac{\phi_\cen-\phi_\upw}{\phi_\dow-\phi_\cen}\,,
    \label{eqnR}
\end{equation}
where $F_\face$ denotes the volumetric face flux and $\left( \cdot \right)_{\face,\text{UD}}$ denotes upwind interpolation.
Contrary to the condition Sweby~\cite{Sweby1984} derived for a flux limiter function $\Psi$, the 
condition in Equation~(\ref{eq:TVDconditionTransferred}) 
does not require the mesh to be uniform. Additionally, the above limiter condition makes use of the whole boundedness region, 
while the original limiter condition introduced in \cite{Sweby1984} only offers a very conservative estimation of the boundedness region.

On unstructured meshes of general topology the computation of $r$ leads to the problem that $\phi_\upw$ is required, the definition of the upwind node, however, is not clear. 
\begin{figure}
    \centering
    \begin{subfigure}[b]{0.49\textwidth}
        \centering
        \includegraphics[width=0.8\textwidth]{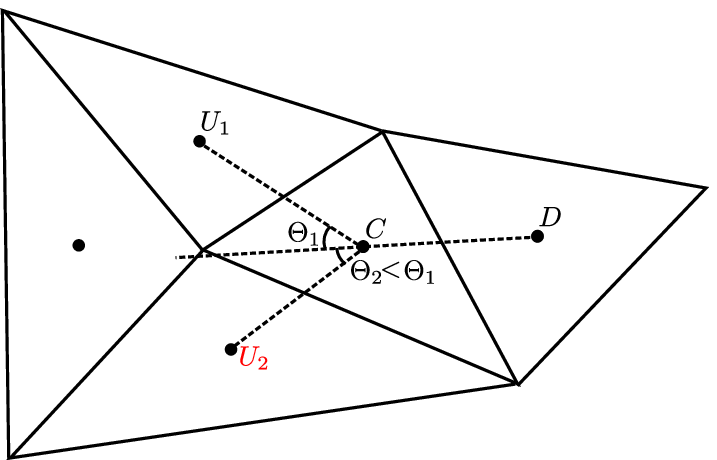}
        \caption{Based on face-neighbour cells only}
        \label{fig:croftVirtualUpwindNode}
    \end{subfigure}
    \begin{subfigure}[b]{0.49\textwidth}
        \centering
        \includegraphics[width=0.8\textwidth]{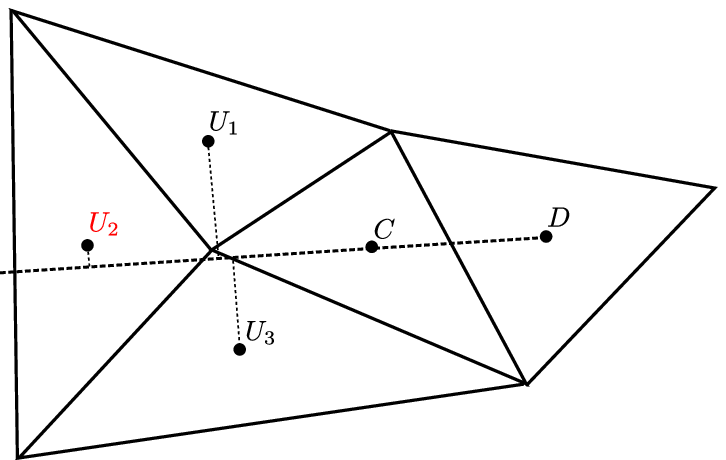}
        \caption{Based on extended stencil}
        \label{fig:altVirtualUpwindNode}
    \end{subfigure}
    \caption{Choice of upwind node on unstructured meshes \cite{Croft1998,Jasak1999}}
    \label{fig:virtualUpwindNode}
\end{figure}
In literature, a variety of different possible remedies have been introduced:
\cite{Croft1998} chooses the upwind cell from all face-neighbouring cells of the centre node
as the one whose centre position is closest to the line through centre and downwind nodes 
(see Figure~\ref{fig:croftVirtualUpwindNode}), whereas \cite{Jasak1999} presents alternatives based 
on an extended stencil (see Figure~\ref{fig:altVirtualUpwindNode}) and compact stencil (see Figure~\ref{figConstrVUnodePos}).
The latter is the most widely utilized approach and introduces a virtual upwind node
as the backward projection of the downwind to centre node.
\begin{figure}
    \centering
    \begin{subfigure}[b]{0.49\textwidth}
        \centering
        \includegraphics[width=0.8\textwidth]{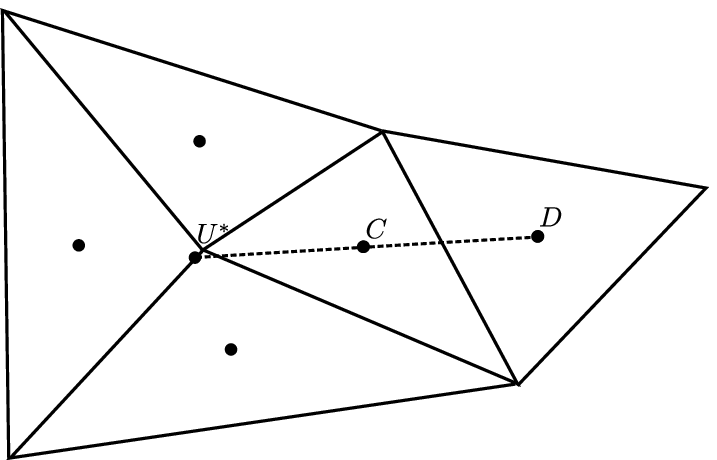}
        \caption{Construction of virtual upwind node position}
        \label{figConstrVUnodePos}
    \end{subfigure}
    \begin{subfigure}[b]{0.49\textwidth}
        \centering
        \includegraphics[width=0.8\textwidth]{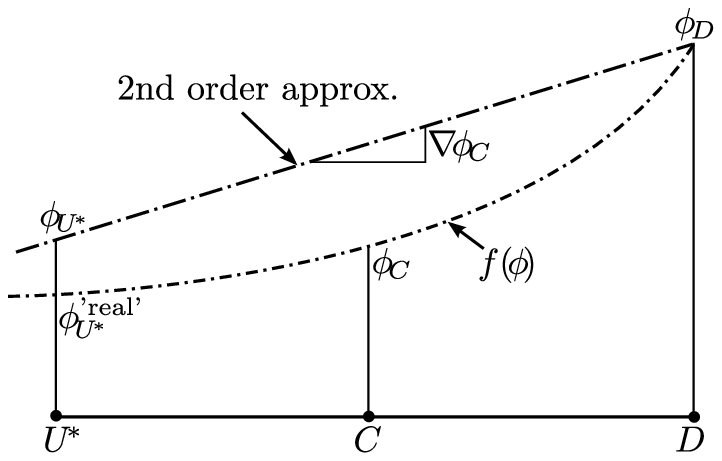}
        \caption{Reconstruction of virtual upwind node value}
        \label{figConstrVUnodeValue}
    \end{subfigure}
    \caption{Virtual upwind node on unstructured meshes based on \cite{Jasak1999,Darwish2003}}
    \label{fig:virtualUpwindNode2}
\end{figure}
The value in the virtual upwind node is then computed by a second-order Taylor series approximation (see Figure \ref{figConstrVUnodeValue})
\begin{equation}
    \phi_{\upw^*} \approx \phi_\dow - 2 \nabla\phi_\cen \cdot \mathbf{d}_{\!\text{CD}} \, .
    \label{eq:VUapprox}
\end{equation}
On uniform hexahedral meshes the exact value for the upwind cell value is recovered,
if the gradient in Equation~(\ref{eq:VUapprox}) is calculated using Gauss' approximation.
Different strategies for calculating the $r$-factor and/or virtual upwind value reported in literature can be found in \cite{Darwish2003,Przulj2001,Li2008,Jasak1999,Hou2012}.
To maintain a bounded solution, different limiting strategies for the virtual upwind have been proposed in literature \cite{Przulj2001,Li2008}.
The simplest strategy is to bound the virtual upwind value by the physical bounds of the transported quantity 
(e.g. between 0 and 1 for the volumetric phase fraction).
\cite{Przulj2001} introduced a local limiting strategy based on the values in the neighbouring (nb) cells
\begin{equation}
    \phi_{U^*}=\text{min}\left(\text{max}\left(\phi_{min}^{nb},\phi_{U^*}\right),\phi^{nb}_{max}\right) \, ,
    \label{eqnLocalLimiting}
\end{equation}
which on unstructured meshes requires an efficient search algorithm to find the computational cell containing the virtual upwind node.
This approach is adopted in the presented work, employing the local vicinity search algorithm of \cite{Loehner1995,Maric2015}.

\subsection{Implicit Mesh-induced Error Correction for Advection}
In order to apply the previously described boundedness criteria
to advection schemes accounting for non-conjunctionality errors, 
the formulation of the correction (Equation~\ref{eqnSkew}) has to be rearranged. 
To this end, an approach has recently been introduced by Denner and vanWachem~\cite{Denner2014,Denner2015}, 
which is based upon including the explicit correction into the interpolation weights.
This is done by extending the general interpolation scheme already defined in Equation~(\ref{eqnInterpolationScheme})
by the correction term and solving the resulting equation for a newly defined interpolation weight $a_\cen$:
\begin{equation}
\begin{aligned}
    \mathbf\phi_\face&=\phi_\cen+\delta_\face \Psi \! \left( r \right)\, \left(\phi_\dow-\phi_\cen\right) +
                   \mathbf m\cdot\nabla\mathbf\phi_{\face'} \\
                 &\stackrel{!}{=} \phi_\cen+a_\cen \, \left(\phi_\dow-\phi_\cen\right)  \\
                 \Rightarrow  a_\cen &= \delta_\face \Psi + \frac{\mathbf m\cdot\nabla\mathbf\phi_{\face'}}{\phi_\dow-\phi_\cen} \, .
    \label{eq:implSkewCoeff} 
\end{aligned}
\end{equation}
To obtain a bounded solution the TVD condition (Equation~\ref{eq:TVDconditionTransferred}) now no longer has to be enforced 
on the flux limiter $\Psi$ of the base scheme, but on the   
resulting interpolation weight $a_\cen$:
\begin{equation}
    0 \leq a_\cen \leq \min\left(r\frac{1-\operatorname{Co}_\face}{\operatorname{Co}_\face}, 1\right)
    \label{eq:TVDconditionTransferredCorrected}
\end{equation}
Thus, in this approach the correction is still explicit but is incorporated directly into the interpolation weights 
which allows for a bounded and formally time-implicit discretisation.

Another problem which must be addressed when aiming at boundedness-preserving non-conjunctionality corrections,
is the coupling between face-interpo\-lation and gradient computation,
which has been noted by numerous researchers \cite{Croft1998,Zhang2006,Zhang2007}.
A face interpolation with non-conjunctionality correction (cf. Equation~\ref{eq:implSkewCoeff})
depends on the face gradient that is interpolated from gradients in the face-neighbouring cell centres, 
which in return may require the face interpolated values to be known (e.g. using Gauss' theorem)
\begin{equation}
    \grad\phi_\cen, \grad\phi_\dow = \frac{1}{|V_P|} \sum\limits_\face \phi_\face \mathbf S_\face \quad \text{and} \quad \grad\phi_{\face'} = f\left( \grad\phi_\cen ,\, \grad\phi_\dow \right) \, .
\end{equation}
To overcome this problem, \cite{Zhang2006} and \cite{Croft1998} introduce different iterative methods, 
requiring the computation of the face interpolated value and face gradient multiple times in a correction loop.
\cite{Zhang2007} introduces an improved approach, based on an implicit calculation of the cell-centred gradient using a least-squares method.

It is shown in this work (Section \ref{ssec:advectionCases}), that the method utilized to compute the gradient 
in the correction term has a major influence on the accuracy of the non-conjunctionality correction for advection terms.
In this work, the least-squares gradient computation introduced in \cite{Zhang2007} and the gradient computation using 
standard Gauss' theorem are utilized and compared in respect of the accuracy of non-conjunctionality corrections. Both gradient computations
are explicit without any correction loops.

\subsection{Implicit Mesh-induced Error Correction for Diffusion}
A straight forward way to account for both mesh-induced errors simultaneously is to consecutively apply the non-conjunctionality 
and non-orthogonality corrections already described. In a first step the 
non-conjunctional corrected gradient of the transported variable $\phi$ analogous to Equation~(\ref{eqnSkew}) is obtained:
\begin{equation}
    \mathbf S_\face\cdot\left(\grad\phi\right)_\face=
    \mathbf S_\face\cdot\left[\left(\grad\phi\right)_{\face'} + \grad\left(\grad\phi\right)_{\face'}\cdot\mathbf m\right]\,. \nonumber
    \label{eqnSnGradNOGrad}
\end{equation}

Using Equation~(\ref{eqnSnGradNO}) to discretise $\mathbf S_\face\cdot\left(\grad\phi\right)_{\face'}$ finally yields:

\begin{eqnarray}
    \mathbf S_f\cdot\left(\grad\phi\right)_\face
    &=&\mathbf \Delta\cdot\left(\grad\phi\right)_{\face'} 
        + \mathbf k\cdot\left(\grad\phi\right)_{\face'} + \mathbf S_\face \cdot 
        \left[ \nabla\left(\grad\phi\right)_{\face'}\cdot\mathbf m \right] \nonumber \\
    &=&|\mathbf\Delta|\frac{\phi_\nei-\phi_\own}{|\mathbf d|} 
        + \mathbf k\cdot\left(\grad\phi\right)_{\face'} + \mathbf S_\face \cdot 
        \left[ \nabla\left(\grad\phi\right)_{\face'}\cdot\mathbf m \right]
    \label{eqnSnGradNOSkew}
\end{eqnarray}

Analogously to the approach previously introduced for advection terms, the above discretisation of the diffusive term can also be rearranged in a way that allows for implicit bounding. This is done by rearranging Equation~(\ref{eqnSnGradNOSkew}) to formally correspond to the uncorrected discretisation of the diffusive term (cf. Equation~\ref{eqnSnGrad}): 
\begin{equation}
  \left(\Gamma_{\!\phi}\right)_{\face}\mathbf S_\face\cdot\left(\grad\phi\right)_\face 
  \stackrel{!}{=}\left(\Gamma_{\!\phi}\right)_{\face,m}|\mathbf S_\face|\frac{\phi_\nei-\phi_\own}{|\mathbf d|}\,.
	\label{eqnConditionDiffusiveImpl}
\end{equation}
Comparing Equation~(\ref{eqnConditionDiffusiveImpl}) with Equation~(\ref{eqnSnGrad}) reveals that the diffusive coefficient $\left(\Gamma_{\!\phi}\right)_{\face,m}$ has to be defined as
\begin{equation}
   \left(\Gamma_{\!\phi}\right)_{\face,m} = 
   \frac{\left(\Gamma_{\!\phi}\right)_{\face}|\mathbf d|}{|\mathbf S_\face|}
   \left[\frac{|\mathbf\Delta|}{|\mathbf d|} + \frac{\mathbf k\cdot\left(\grad\phi\right)_{\face'}}{\phi_\nei-\phi_\own} + \frac{\mathbf S_\face \cdot \left[ \nabla\left(\nabla\phi\right)_{\face'}\cdot\mathbf m \right]}{\phi_\nei-\phi_\own}\right] \, .
    \label{eqnSnGradNOSkewImplicit}
\end{equation}

Theoretically this formulation now enables the implicit application of 
limiting criteria to the diffusive coefficient $\left(\Gamma_{\!\phi}\right)_{\face,m}$, 
which in turn contains the explicit correction. However, until now it remains uncertain as to whether limiting is needed at all 
and if so, the criterion required is missing. Both issues will be addressed in the results section. 

\section{Utilized Finite Volume Discretisation}
\label{sec:discretization}
The developed correction strategies are applied to the discretised phase fraction advection equation, i.e.~the VoF transport eqn.
\begin{equation}
    \partial_t\alpha+\nabla\cdot\left(\alpha \mathbf{u}\right) = 0
    \label{eqnAlpha}
\end{equation}                                        
and the discretised species transport equation underlying the CST model \cite{Deising2016}. 
Employing a harmonic mean diffusion coefficient, the simplest version of the latter can be written as (cf.~\cite{Deising2016})
\begin{align}
    &\partial_t c +\nabla\cdot\left(c \mathbf{u}\right)
    = \nabla\cdot\left({\langle \text{D} \rangle_{\!\text{h}}}\grad c\right)
     +\nabla\cdot\left({{\langle \text{D} \rangle_{\!\text{h}}} \text{K} c }\grad\alpha\right) \, , \label{eqnCST} \\
     &\text{with K}= \frac{\operatorname{H}-1}{1+\alpha\left(\operatorname{H}-1\right)} \text{ and } \operatorname{H} \text{ the Henry coefficient} \nonumber .
\end{align}
The above equations can be discretised in many different ways to account for mesh-induced errors.
Suitable candidates -- based on the selection in the previous Sections -- 
are identified in the remainder of this Section and tested in Section \ref{sec:results}.
The discretised form of the phase fraction transport equation, using time-centered Crank-Nicolson time discretisation, reads
\begin{equation}
    \alpha_{\cen}^n = \alpha_{\cen}^o - \frac{\Delta t^n}{|V|} \sum\limits_\face \frac{1}{2} \left( \alpha_{\face}^o + \alpha_{\face}^n \right) F_{\face}^o \, , 
    \text{ with } F_{\face}^o = \mathbf{u}_{\face}^o \cdot \mathbf{S}_{\face} \, .
\end{equation}
The most crucial decision in discretising the advection term is how to compute the face-interpolated 
value $\alpha_{\face}$. This work considers four different schemes, namely the Upwind \scheme{UDS}, Central Differences (\scheme{CDS}), 
\scheme{CICSAM} \cite{Ubbink1997} and a specific local blended scheme (\scheme{LB}). 
The employed \scheme{LB} scheme is similar to the \scheme{CICSAM} scheme in that it uses the \scheme{CICSAM} scheme's blending function 
to blend between the central differencing and downwind scheme,
resulting in a formally unbounded interpolation scheme.
In the \scheme{LB} case, boundedness will be enforced by implicit correction once, while in the \scheme{CICSAM} case,
boundedness is enforced twice: by the non-linear limiter of the scheme and by limiting the skewness corrected face interpolation.

Each of the schemes is tested without correction (\scheme{UC}), with standard explicit correction 
(\scheme{EC}) and with the newly proposed correction, from
now on referred to as Semi-Implicit Skewness Correction (\scheme{SISC}) scheme. Both corrections make use of the gradient operator
which is discretised by both, Gaussian Gradient (\scheme{GG}) and Least Squares Fit (\scheme{LSF}). 
An overview of all interpolation schemes, correction approaches and gradient calculation methods 
varied against each other for discretisation of the advection term is given in Table~\ref{tabDiscretisationCombinationAdvection}.
\begin{table}[htb]
    \centering
    \caption{Overview of all methods used to discretise the face-interpolated value $\alpha_f$.}
    \label{tabDiscretisationCombinationAdvection}
    \begin{tabular}{@{}lll@{}}
    \toprule
    Interpolation Scheme  & Correction Approach & Gradient Calculation  \\ \midrule
    \scheme{UDS}          & \scheme{UC}         & \scheme{GG}           \\
    \scheme{CDS}          & \scheme{EC}         & \scheme{LSF}          \\
    \scheme{CICSAM}       & \scheme{SISC}       &                       \\
    \scheme{LB}           &                     &                       \\ \bottomrule
\end{tabular}
\end{table}

In this work, we test different gradient schemes for calculation of the correction part
and utilize the least-squares gradient to compute the virtual upwind value on non-uniform meshes.
It is emphasised that on uniform Cartesian meshes the Gaussian gradient 
should be utilized for the virtual upwind calculation to maintain accuracy.

It is crucial to discretise the left hand side of the species transport equation~(\ref{eqnCST}) identical
to the phase fraction transport equation in order to avoid artificial species transfer,
i.e. identical discretisation schemes for the respective terms and corrections need to be employed.
A suitable discretisation of the right hand side as utilized in the presented work reads
\begin{align}
    & \sum\limits_\face {\langle \text{D} \rangle_{\!\text{h,f}}^n} \left[\mathbf S_{\face} \cdot \left(\grad c \right)_{\face}^n \right]
    - \sum\limits_\face {\langle \text{D} \rangle_{\!\text{h,f}}^n} \left( \text{K}c \right)_{\face}^n  
    \left[\mathbf S_{\face} \cdot \left(\grad \alpha \right)_{\face}^n \right] \, .
\end{align}
Since the phase fraction transport equation is solved first, the phase fraction field $\alpha^n$ at the new time level is already explicitly available,
allowing for the fully time implicit discretisation shown above. 
The gradient in face normal direction in both terms need to be discretised identically, whereby in the scope of this work 
three different discretisation methods are tested. Firstly, without 
any correction, which therefore corresponds to Equation~(\ref{eqnSnGrad}). 
Secondly, by applying the explicit non-orthogonal correction as   
proposed by Jasak~\cite{Jasak1996}, cf. Equation (\ref{eqnSnGradNO}), and thirdly, by additionally accounting for 
non-conjunctionality errors as formulated in (\ref{eqnSnGradNOSkewImplicit}). These three variants are from now on referred
to as \scheme{UC}, \scheme{NO} and \scheme{NO/NC}, respectively.
\begin{table}[htb]
    \centering
    \caption{Overview of all used methods varied against each other to discretise the diffusive terms of the CST model.}
    \label{tabDiscretisationCST}
    \begin{subtable}[h]{0.17\textwidth}
      \begin{tabular}{l}
      \Xhline{2\arrayrulewidth}
      $\mathbf S_{\face} \cdot \left(\grad \phi \right)_{\face}$ \\[5pt]
       \hline \\[-6pt]
       \scheme{UC} \\
       \scheme{NO} \\
       \scheme{NO}/\scheme{NC} \\
      \Xhline{2\arrayrulewidth}
      \end{tabular}
\caption{Laplacian}
\label{tab:discrCSTa}
    \end{subtable}
    \begin{subtable}[h]{0.74\textwidth}
      \begin{tabular}{lll}
      \Xhline{2\arrayrulewidth}
      \multicolumn{3}{c}{$\langle \text{D} \rangle_{\!\text{h,f}}^n$,
                        $\left( \text{K}c \right)_{\face}$}    \\[5pt] \hline
         Interpolation Scheme & Correction Appr. & Gradient Calc. \\[5pt] \hline
         \scheme{CDS}         & \scheme{UC}    & \scheme{GG}  \\
                              & \scheme{EC}    &              \\
        \Xhline{2\arrayrulewidth}
      \end{tabular}
\caption{Diffusive flux}
\label{tab:discrCSTb}
    \end{subtable}
\hfill
\end{table}
The diffusion coefficient $\langle \text{D} \rangle_{\!\text{h,f}}^n$ and the term $\left(\text{K}c\right)_f$ 
are linearly interpolated, once with and once without non-conjunctionality correction. 
Table~\ref{tabDiscretisationCST} summarises all 
used discretisations for the different terms of the CST model.

\section{Results and Discussion}
\label{sec:results}
\vspace{-2pt}
To evaluate the performance of all discretisation methods presented in Section~\ref{sec:discretization}, 
including the newly developed correction strategies \scheme{SISC} and \scheme{NO/NC}, they
are verified using a series of test cases involving steep gradients and discontinuities.
Two aspects are considered to judge the discretisation performance: 
the obtained solution should accurately approximate the analytical solution
and at the same time boundedness must be preserved (as required for both phase fraction and species concentration).

The evaluation comprises of two steps.
First, the discretisation of the phase fraction transport equation is varied according 
to Table~\ref{tabDiscretisationCombinationAdvection} and verified.
Then secondly, the diffusive terms in the CST-model are discretised with different variants as given in Table~\ref{tabDiscretisationCST}. 
In both steps,the respective discretization performance is verified via different test cases.
\begin{figure}[htb]
    \centering
    \includegraphics[width=0.30\textwidth]{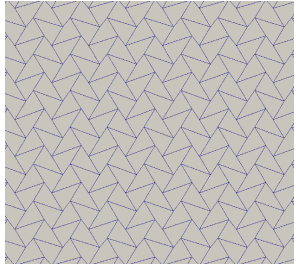}
    \caption{Structure of distorted mesh comprising of non-orthogonal and non-conjunctionality errors.}
    \label{figMesh}
\end{figure}
In all test cases 2D meshes are used. To provoke mesh induced errors the mesh is systematically distorted in such a way that non-orthogonality
as well as non-conjunctionality errors are present. Figure \ref{figMesh} shows the mesh structure which is identical for all cases.

\subsection{Correction of VoF-based phase fraction advection term}
\label{ssec:advectionCases}
\subsubsection{Test Cases}
First we consider a simple 2D plug flow, whereby the transported field is 
advected from left to right by a constant velocity field (see Figure~\ref{figAdvectionSketchPlug}). 
The second case examines the shape-preserving properties of different advection schemes 
when a circular shape is translated through a distorted mesh over a distance equal to
12 times its diameter (see Figure~\ref{figAdvectionSketchCircle}). 
Here, the utilised average mesh resolution is approximately 15 cells per diameter of the circle. 
The numerical time step in both simulations is set in such a way that the Courant number limit 
$\text{Co} = |\mathbf{u}|\Delta t / \Delta x  \le 0.1$ is maintained. 

\begin{figure}[htb]
    \centering
    \begin{subfigure}[b]{0.48\textwidth}
        \begin{overpic}[width=\textwidth]{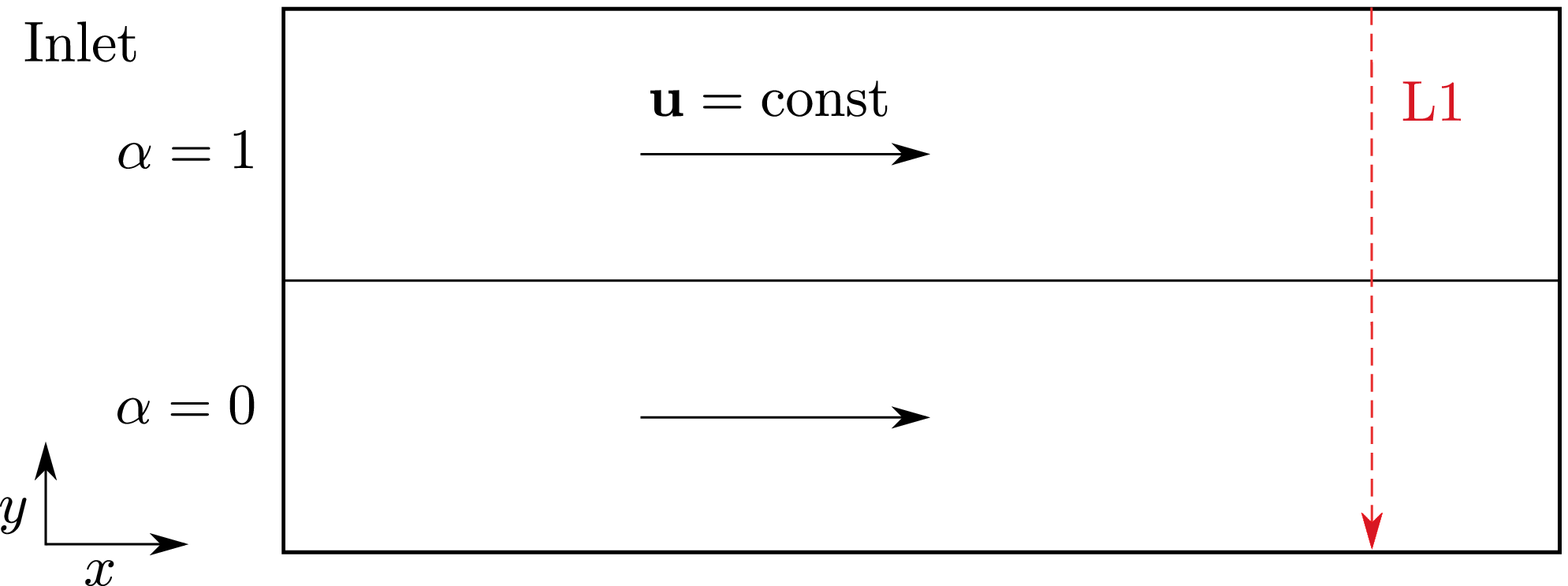}
        \end{overpic}
        \caption{Plug flow.}
        \label{figAdvectionSketchPlug}
    \end{subfigure}
    \begin{subfigure}[b]{0.01\textwidth}
        \includegraphics[width=\textwidth]{figures/blank.eps}
    \end{subfigure}
    \begin{subfigure}[b]{0.48\textwidth}
        \begin{overpic}[width=\textwidth]{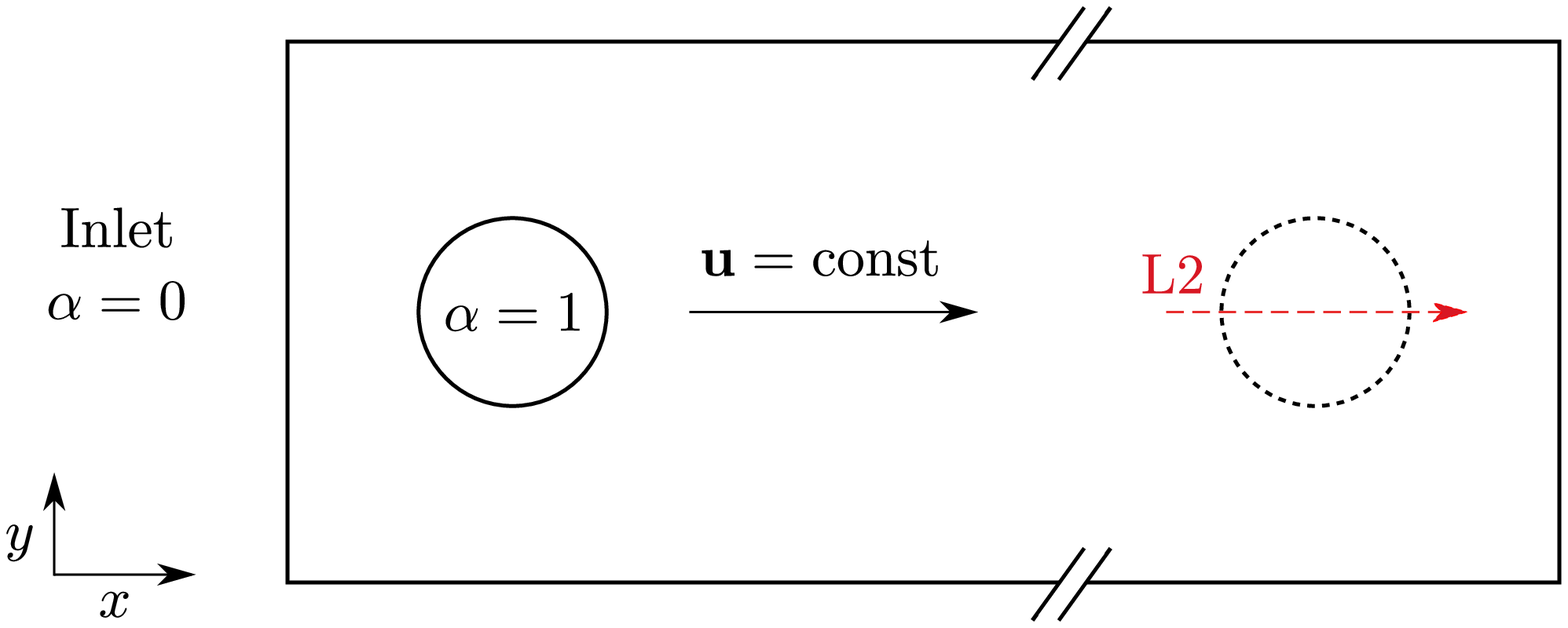}
        \end{overpic}
        \caption{Translation of circular shape.}
        \label{figAdvectionSketchCircle}
    \end{subfigure}
    \caption{Set-up of test cases to evaluate different discretisations of the advection term. The dashed red lines indicate the
             paths over which $\alpha$ is plotted for evaluation reasons.}
    \label{figAdvectionSketch}
\end{figure}

\subsubsection{Results}
Results are visualized by plotting the phase fraction profile over the path `L1' and `L2' respectively 
(see Figure \ref{figAdvectionSketch}).
Without any correction all interpolation schemes show poor results in both test cases. As might be expected the usage of 
\scheme{CDS} or \scheme{LB} generates strongly unbounded, inaccurate solutions in all cases. While boundedness is ensured by the use
of \scheme{UDS} or \scheme{CICSAM} interpolation, the accuracy is insufficient. More specifically, high diffusion is 
observed when applying \scheme{UDS} and severe deformation is ovserved with \scheme{CICSAM}. The respective plots are shown in
Figure~\ref{figAdvectionNoCorr}, where it should be noted that some lines cannot be detected as the unboundedness of the solution
is too severe.
\\
\\
\begin{figure}[htb]
    \centering
    \graphicspath{{gnuplot/}}
    \resizebox{1.0\textwidth}{!}{\input{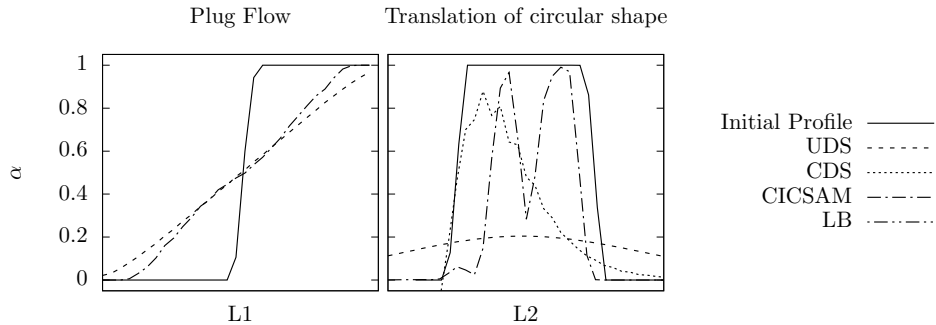}}
    \caption{Phase fraction profiles for different advection schemes without any correction.}
    \label{figAdvectionNoCorr} 
\end{figure}

Adding explicit correction worsens the results even further. Except for the \scheme{UDS} scheme 
all generated solutions are unbounded and all of them suffer heavily from a lack of accuracy. 
This is the case irrespective of whether the gradient is calculated using \scheme{GG} or \scheme{LSF}, where 
applying the latter still produces the better results.

Now, evaluating the \scheme{SISC} scheme, we can state that for all tested combinations boundedness is preserved up to
$10^{-6}$ (see Figure~\ref{figAdvectionSISC}). Another interesting finding is that the \scheme{LSF} of the gradient outperforms the Gaussian gradient
in all cases. Except for the \scheme{UDS} scheme, where the strong diffusion overrides all other effects, the difference 
is quite significant. 
Using the \scheme{SISC}-bounded \scheme{CDS}, diffusion has a huge effect and smears out the initial step profiles. Applying \scheme{CICSAM}
yields a significant improvement because it is better able to preserve steep gradients. 
The best result, however, is achieved by the \scheme{LB} scheme, where almost the exact shape is preserved when using \scheme{LSF} for gradient computation. 
\begin{figure}[htb]
    \centering
    \graphicspath{{gnuplot/}}
    \resizebox{1.0\textwidth}{!}{\input{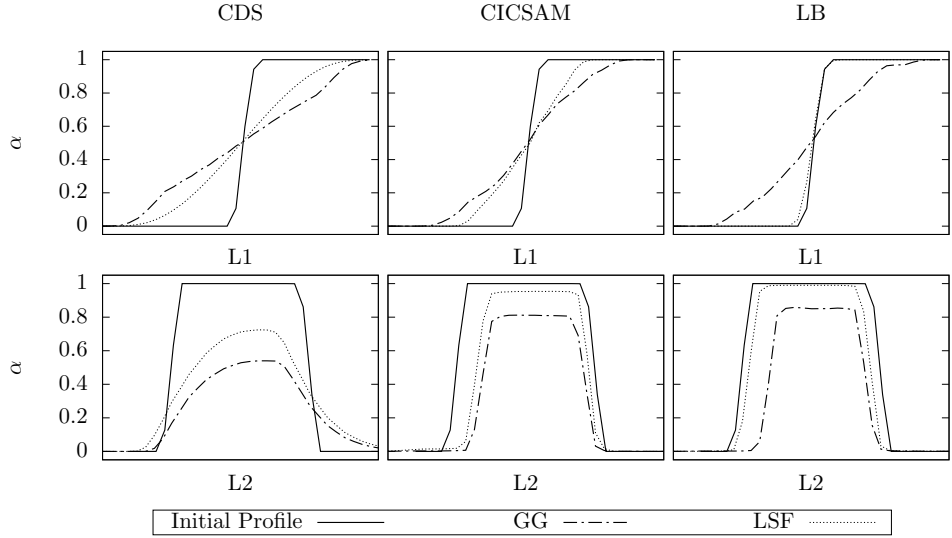}}
    \caption{Phase fraction profiles using the \scheme{SISC} scheme in combination with different base schemes and  gradient computations.}
    \label{figAdvectionSISC} 
\end{figure}

To finally assess all relevant scheme combinations the phase fraction field is also visualized 
(see Figures~\ref{figAdvectionVisual} and \ref{figAdvectionSISCVisual}). Set-ups not considered
as relevant are those which lead to unbounded results and those using the Gaussian gradient approximation because \scheme{LSF}
performs equally or better in all cases. Other than the results gained with the \scheme{SISC} scheme this leaves only the non-corrected
\scheme{UDS} and \scheme{CICSAM} schemes, and for the explicitly corrected set-ups solely the \scheme{UDS} scheme. 
\begin{figure}[htb]                                                                              
	\centering
	\begin{subfigure}[b]{0.21\textwidth}
		\centering
		\includegraphics[width=\textwidth]{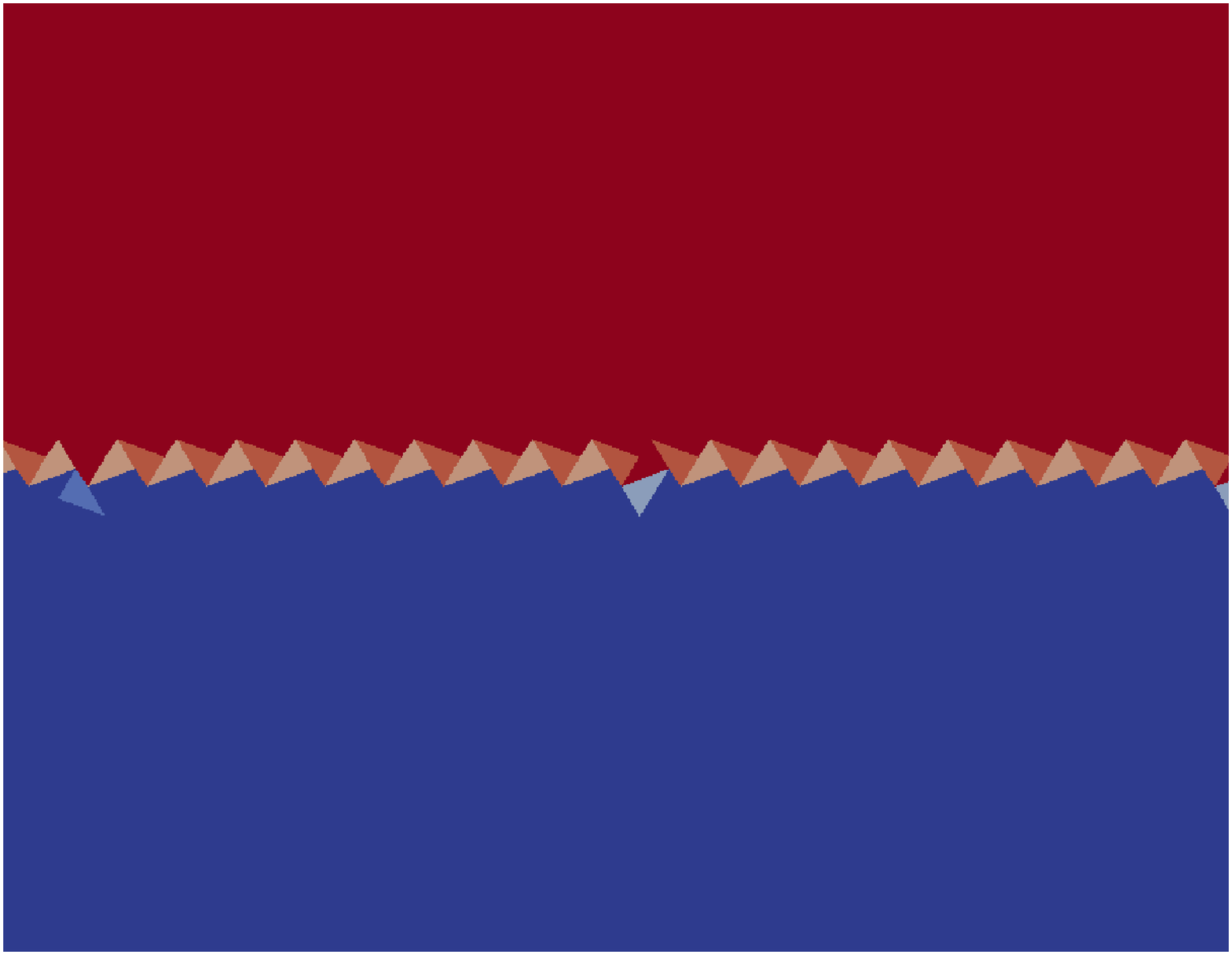}
	\end{subfigure}
	\begin{subfigure}[b]{0.21\textwidth}
		\centering
		\includegraphics[width=\textwidth]{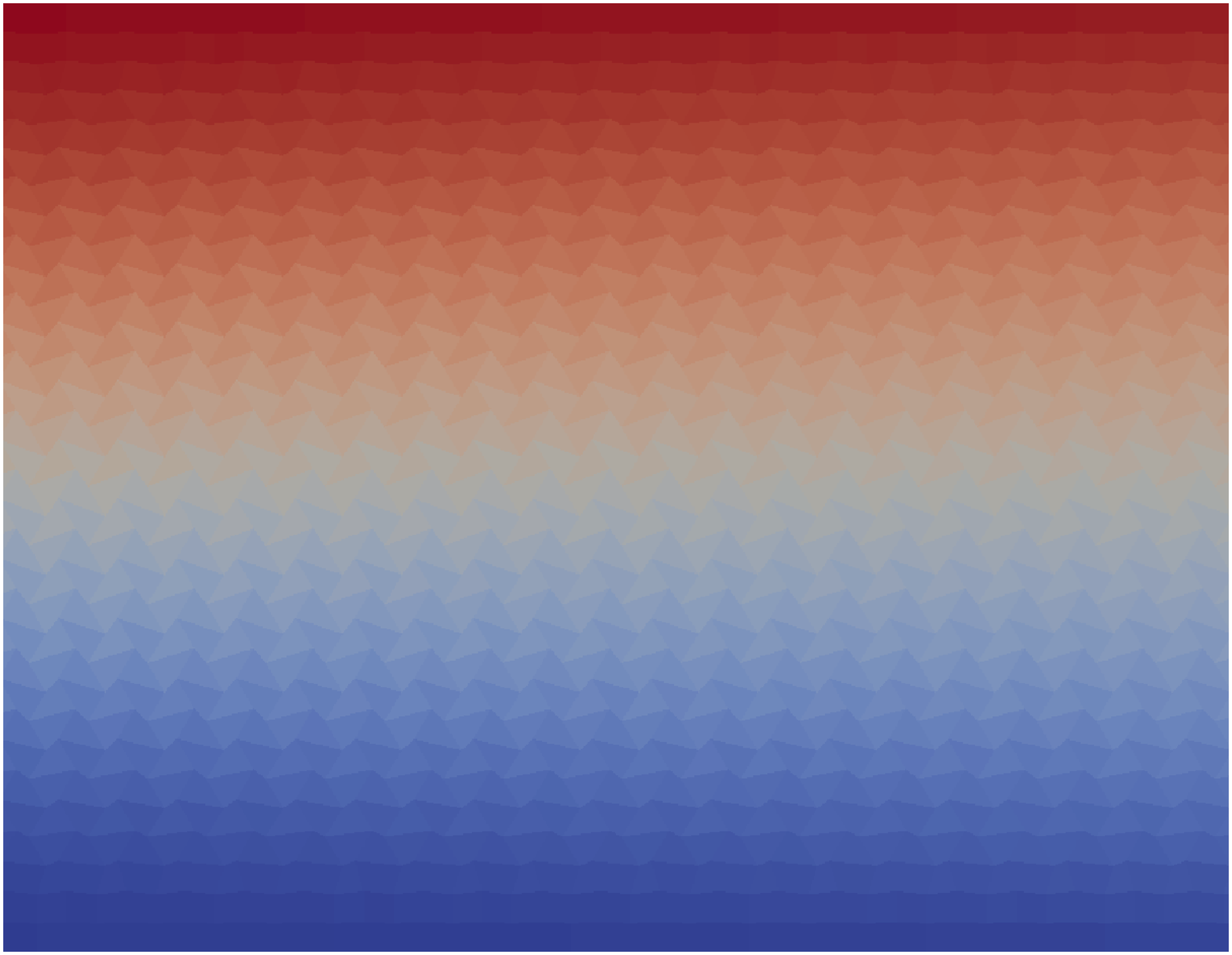}
	\end{subfigure}
	\begin{subfigure}[b]{0.21\textwidth}
		\centering
		\includegraphics[width=\textwidth]{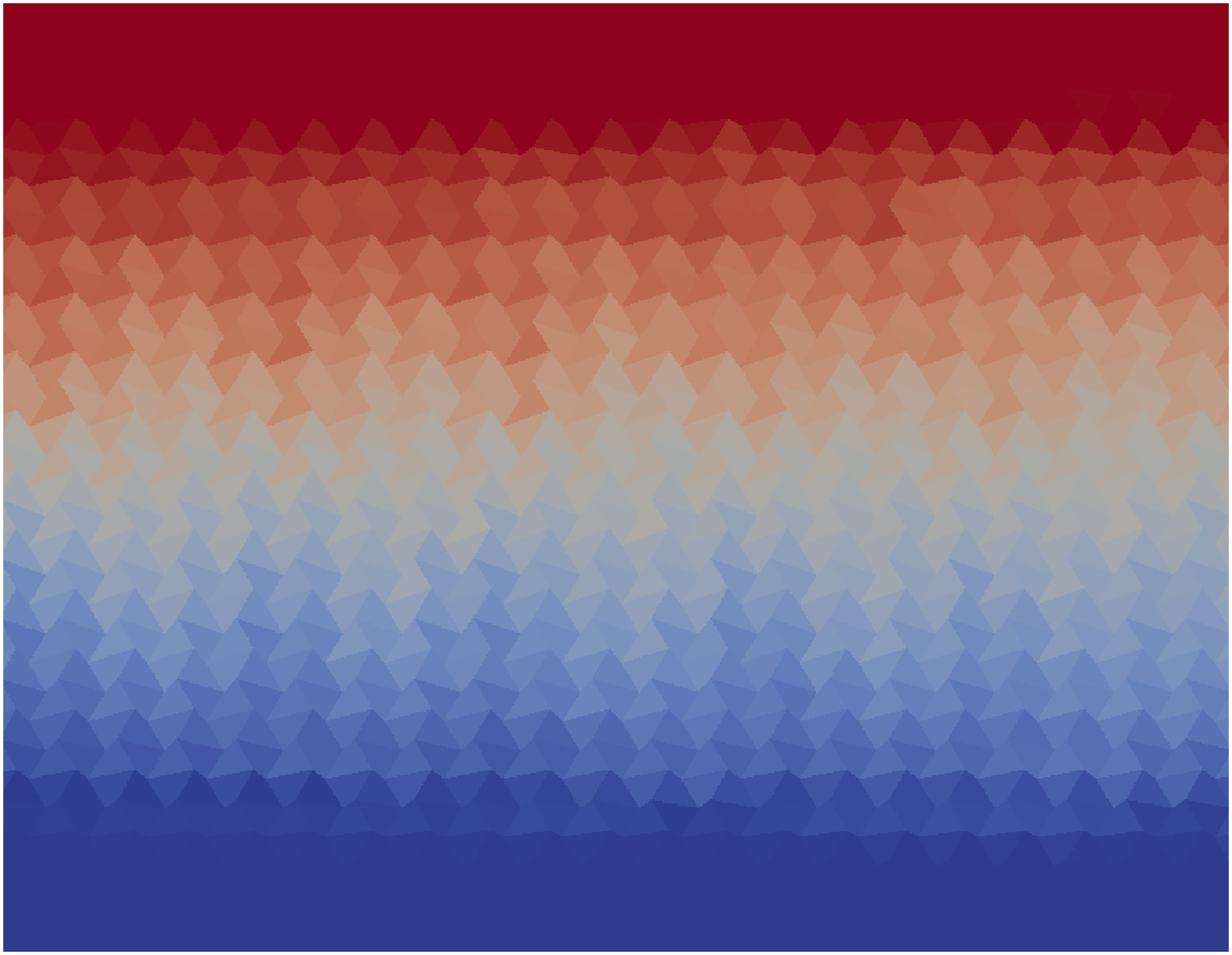}
	\end{subfigure}
	\begin{subfigure}[b]{0.21\textwidth}
		\centering
		\includegraphics[width=\textwidth]{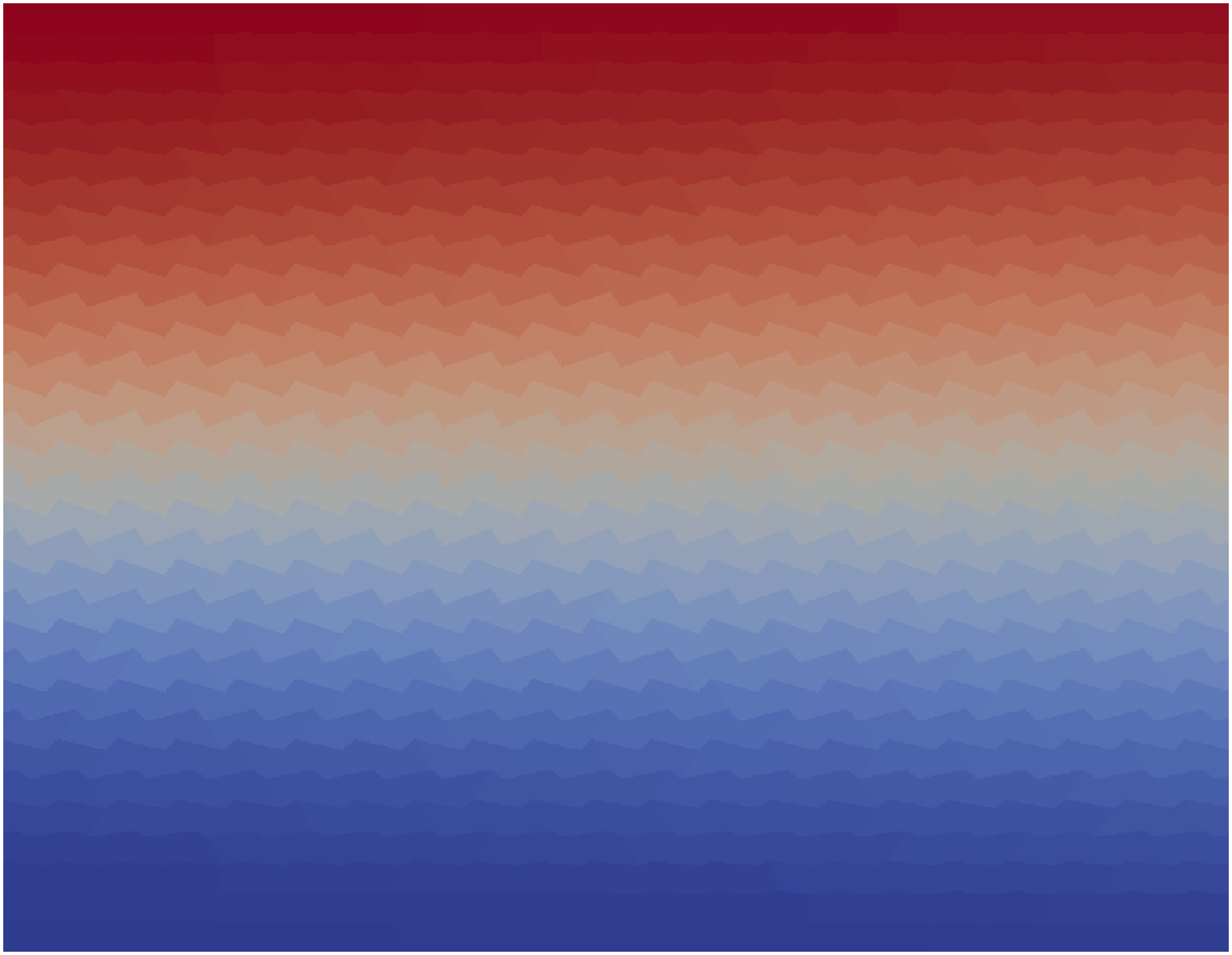}
	\end{subfigure}
	\begin{subfigure}[b]{0.21\textwidth}
		\centering
		\includegraphics[width=\textwidth]{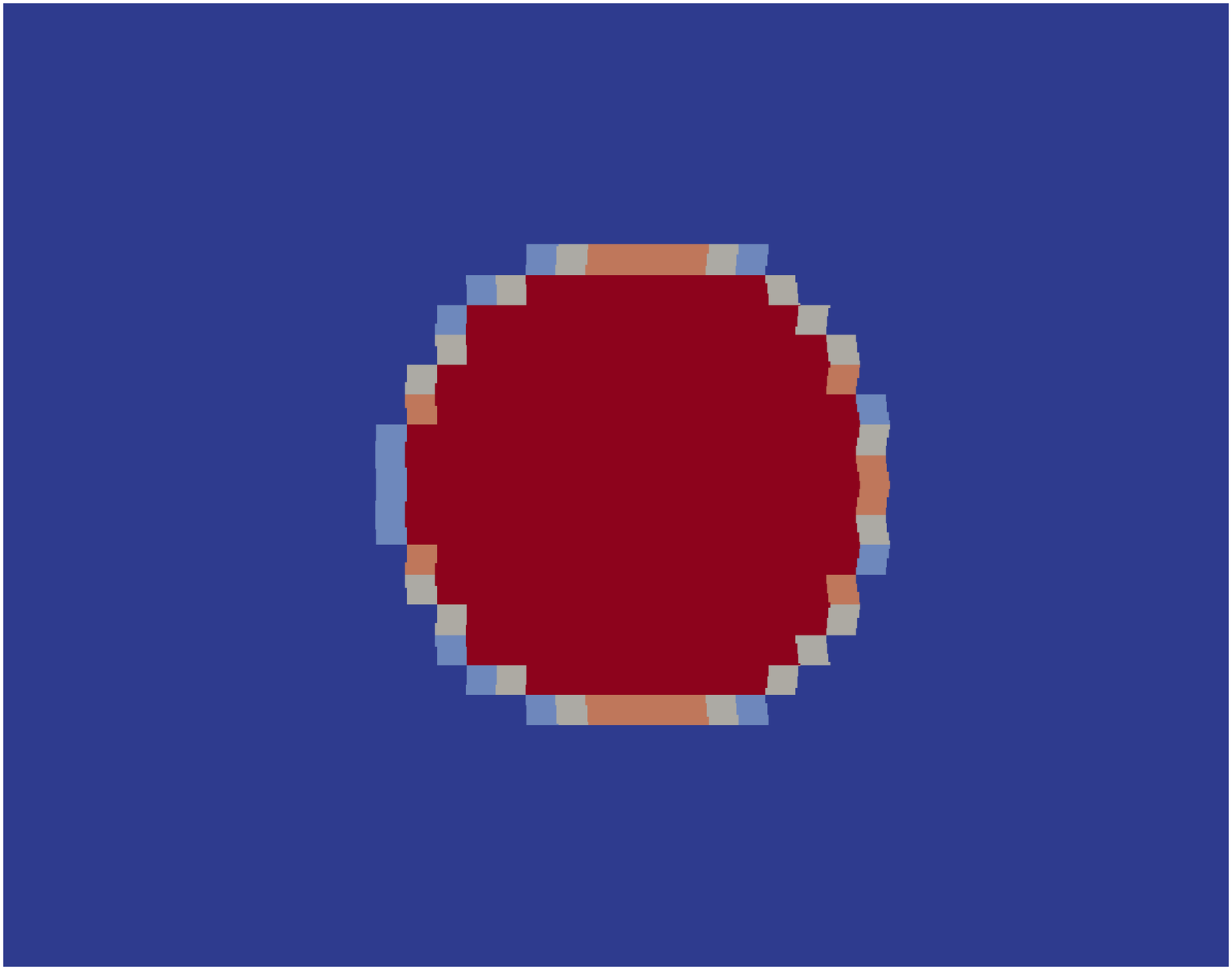}
	\caption*{Initial state}
	\end{subfigure}
	\begin{subfigure}[b]{0.21\textwidth}
		\centering
		\includegraphics[width=\textwidth]{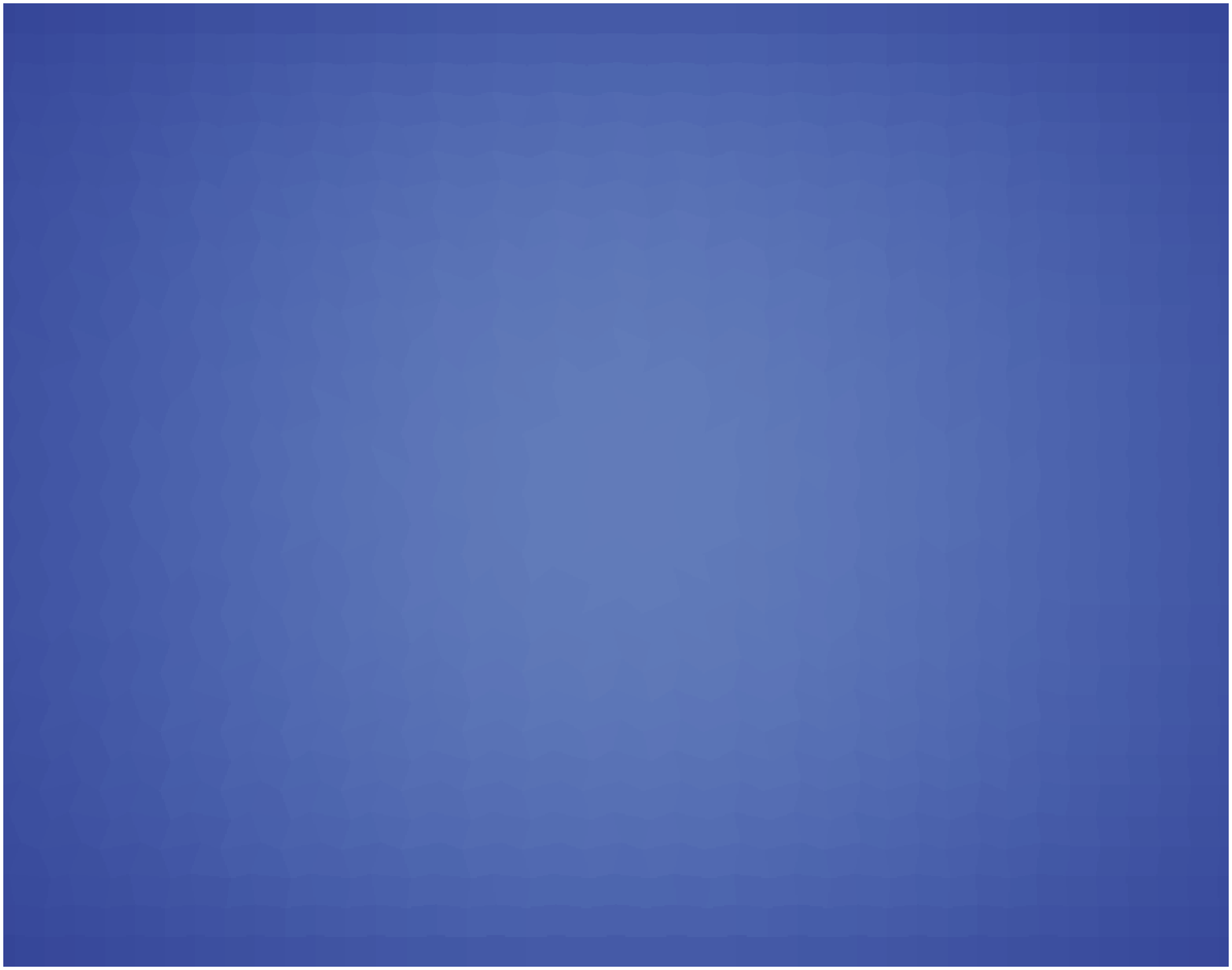}
	\caption*{UDS\,/\,UC}
	\end{subfigure}
	\begin{subfigure}[b]{0.21\textwidth}
		\centering
		\includegraphics[width=\textwidth]{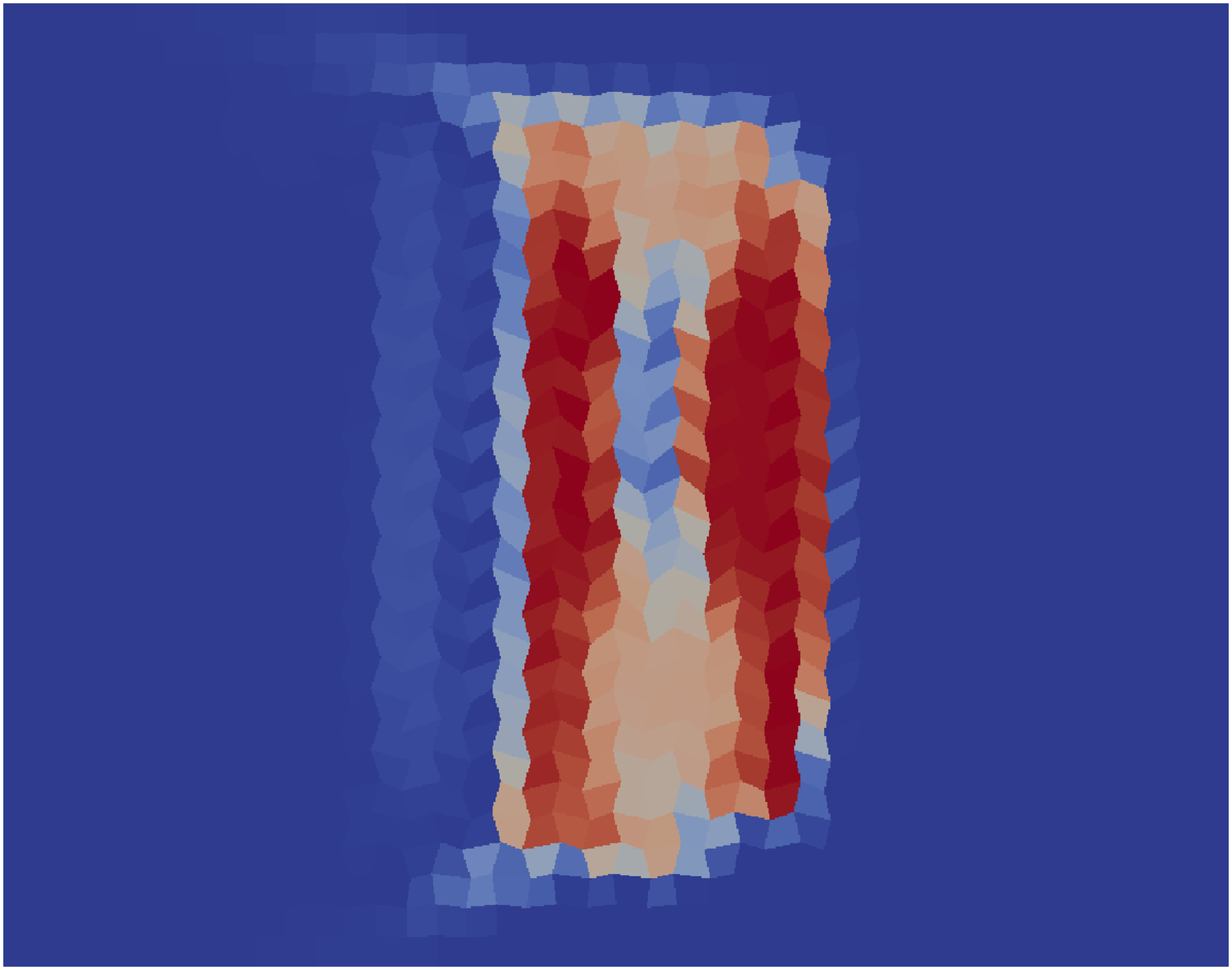}
	\caption*{CICSAM\,/\,UC}
	\end{subfigure}
	\begin{subfigure}[b]{0.21\textwidth}
		\centering
		\includegraphics[width=\textwidth]{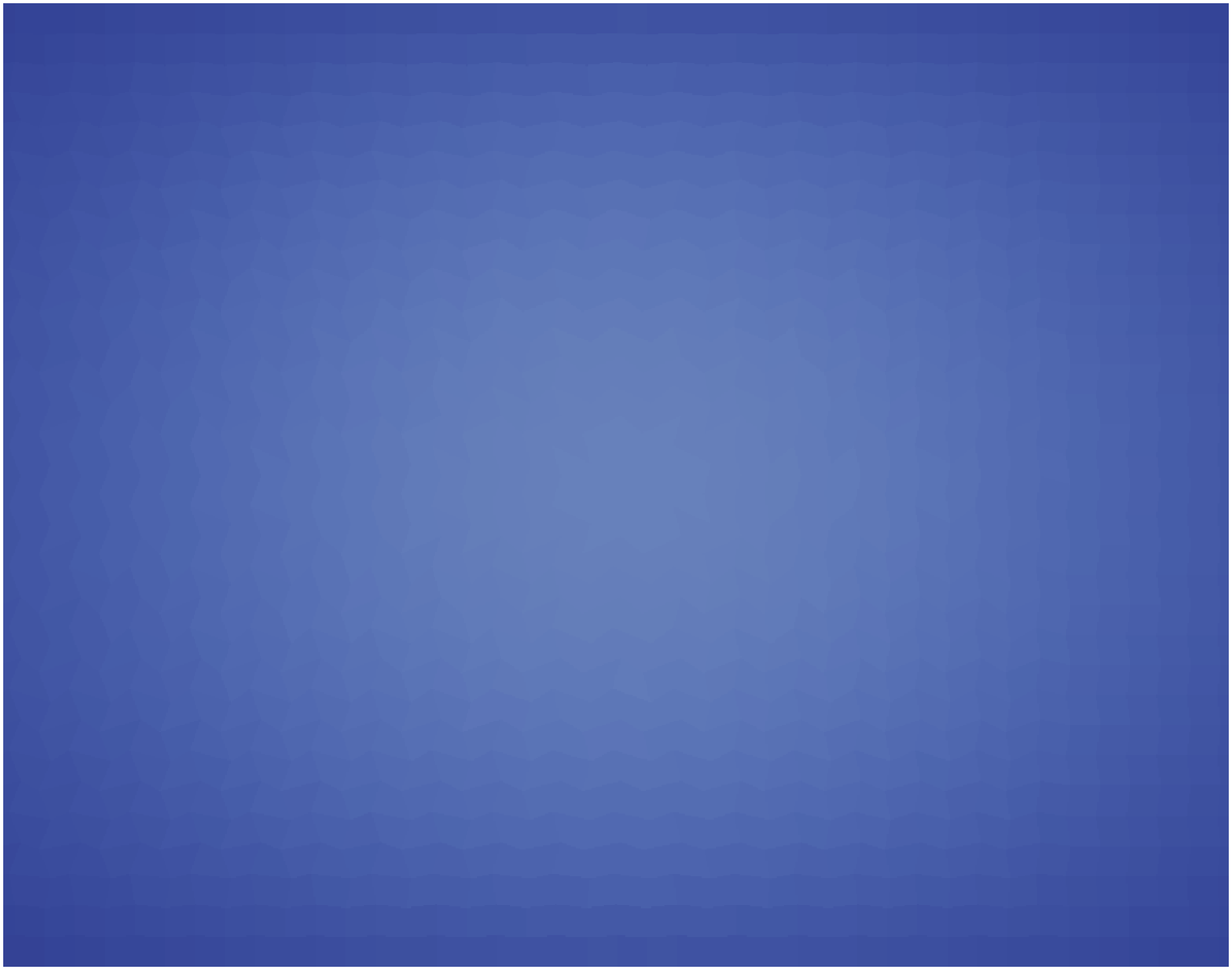}
	\caption*{UDS\,/\,EC}
	\end{subfigure}
    \caption{Visualisation of phase fraction field for all boundedness preserving scheme combinations excluding \scheme{SISC}.}
	\label{figAdvectionVisual}
\end{figure}                                 

Figure \ref{figAdvectionVisual} 
depicts the fields excluding \scheme{SISC}. The \scheme{UDS} scheme without 
and also with explicit correction is dominated by diffusive effects
which leads to a smeared out $\alpha$-distribution. \scheme{CICSAM} results in a slightly sharper solution, which, however, still can not be
considered as satisfactory. It can therefore be concluded that none of the commonly used advection schemes meet the requirements concerning
accuracy and boundedness.
\begin{figure}[htb]
	\centering
	\begin{subfigure}[b]{0.19\textwidth}
		\centering
		\includegraphics[width=\textwidth]{figures/plugInitial.eps}
	\end{subfigure}
	\begin{subfigure}[b]{0.19\textwidth}
		\centering
		\includegraphics[width=\textwidth]{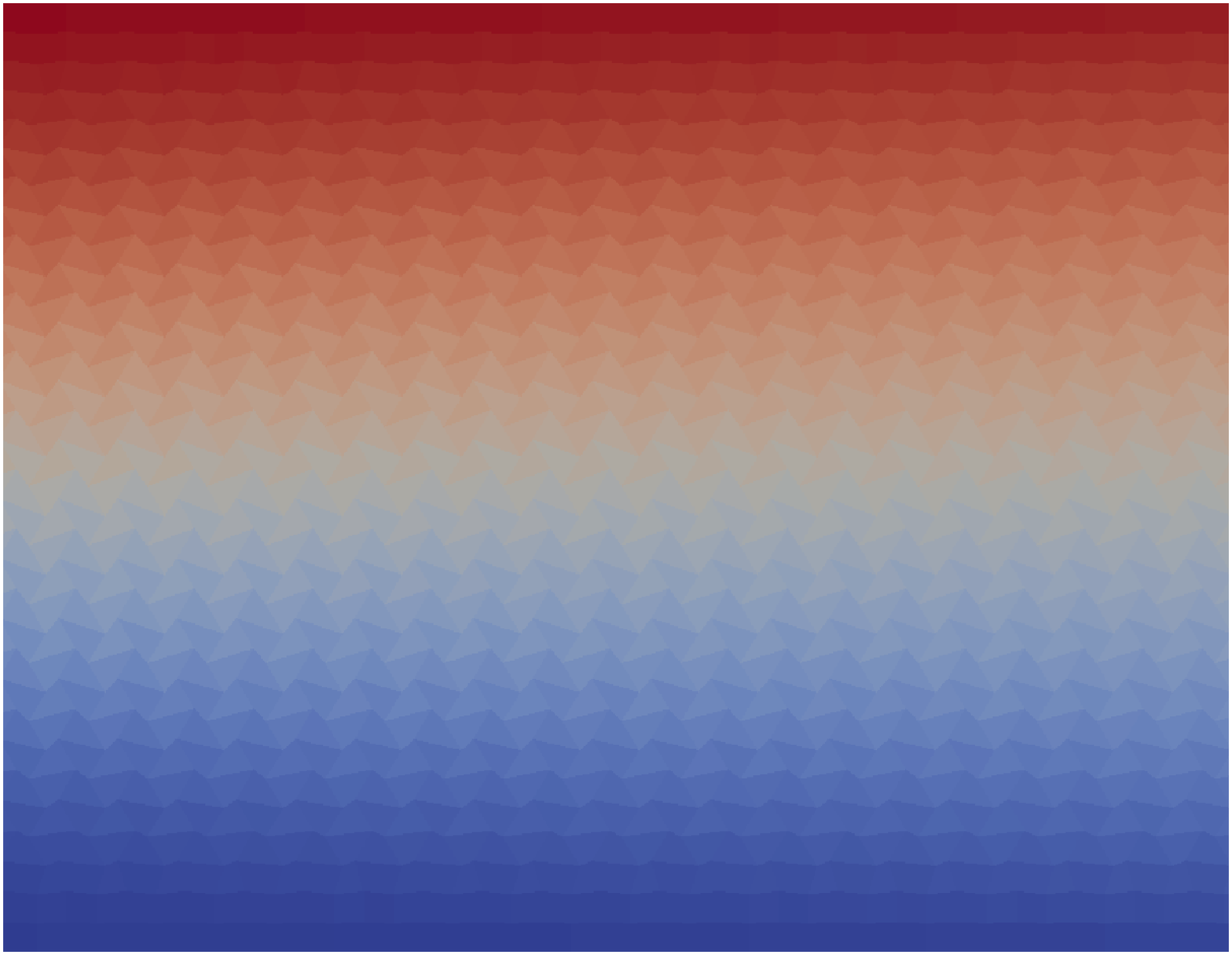}
	\end{subfigure}
	\begin{subfigure}[b]{0.19\textwidth}
		\centering
		\includegraphics[width=\textwidth]{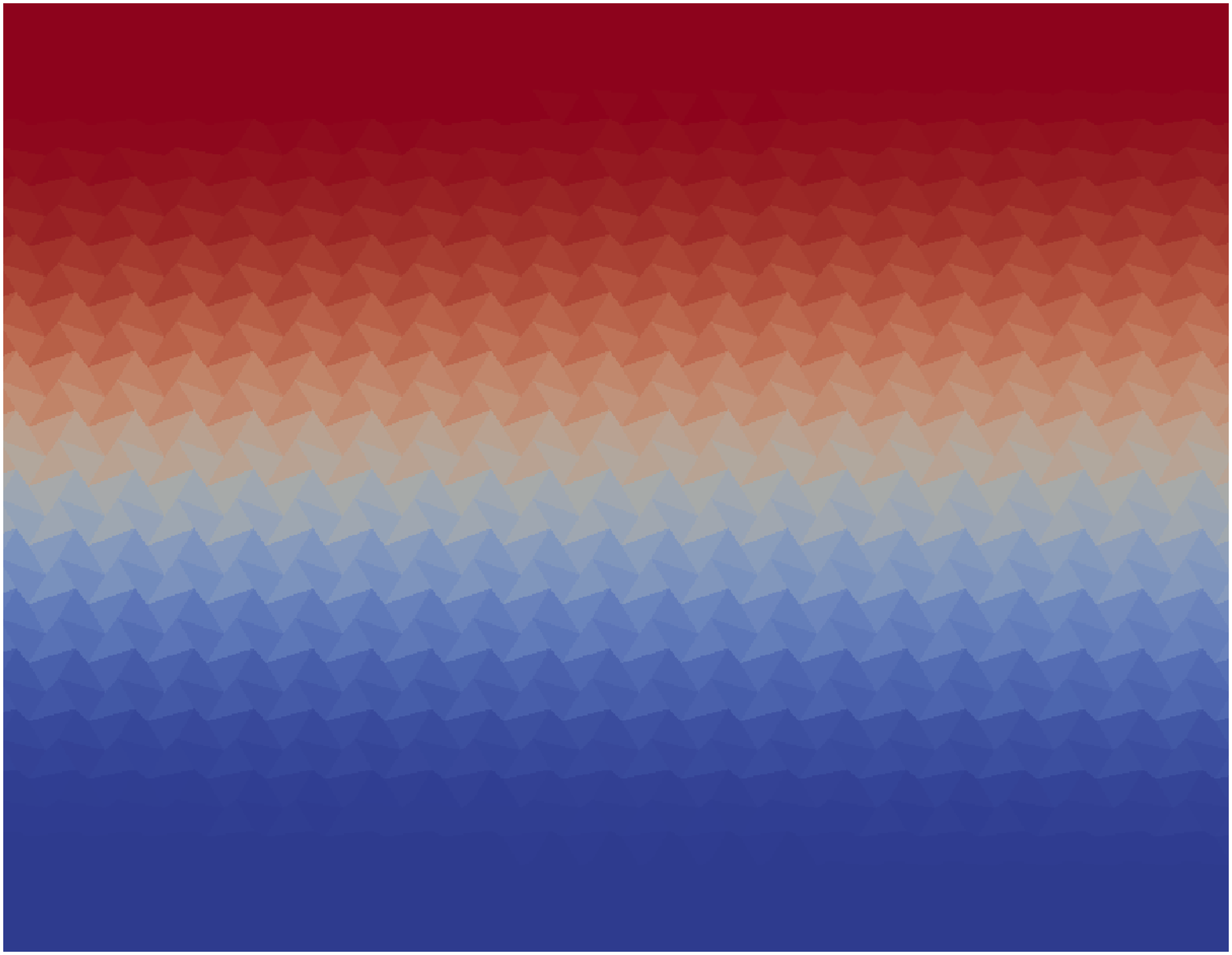}
	\end{subfigure}
	\begin{subfigure}[b]{0.19\textwidth}
		\centering
		\includegraphics[width=\textwidth]{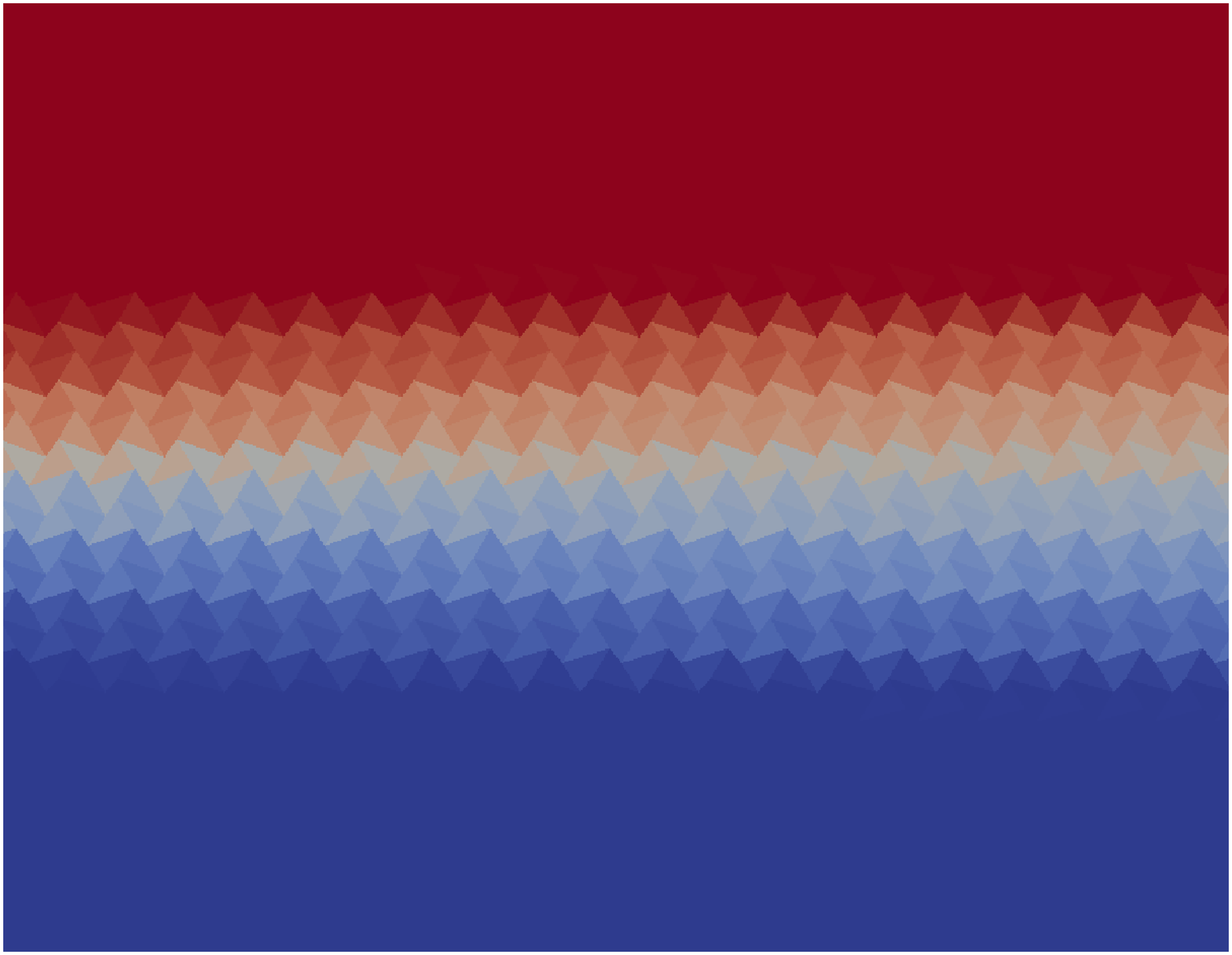}
	\end{subfigure}
	\begin{subfigure}[b]{0.19\textwidth}
		\centering
		\includegraphics[width=\textwidth]{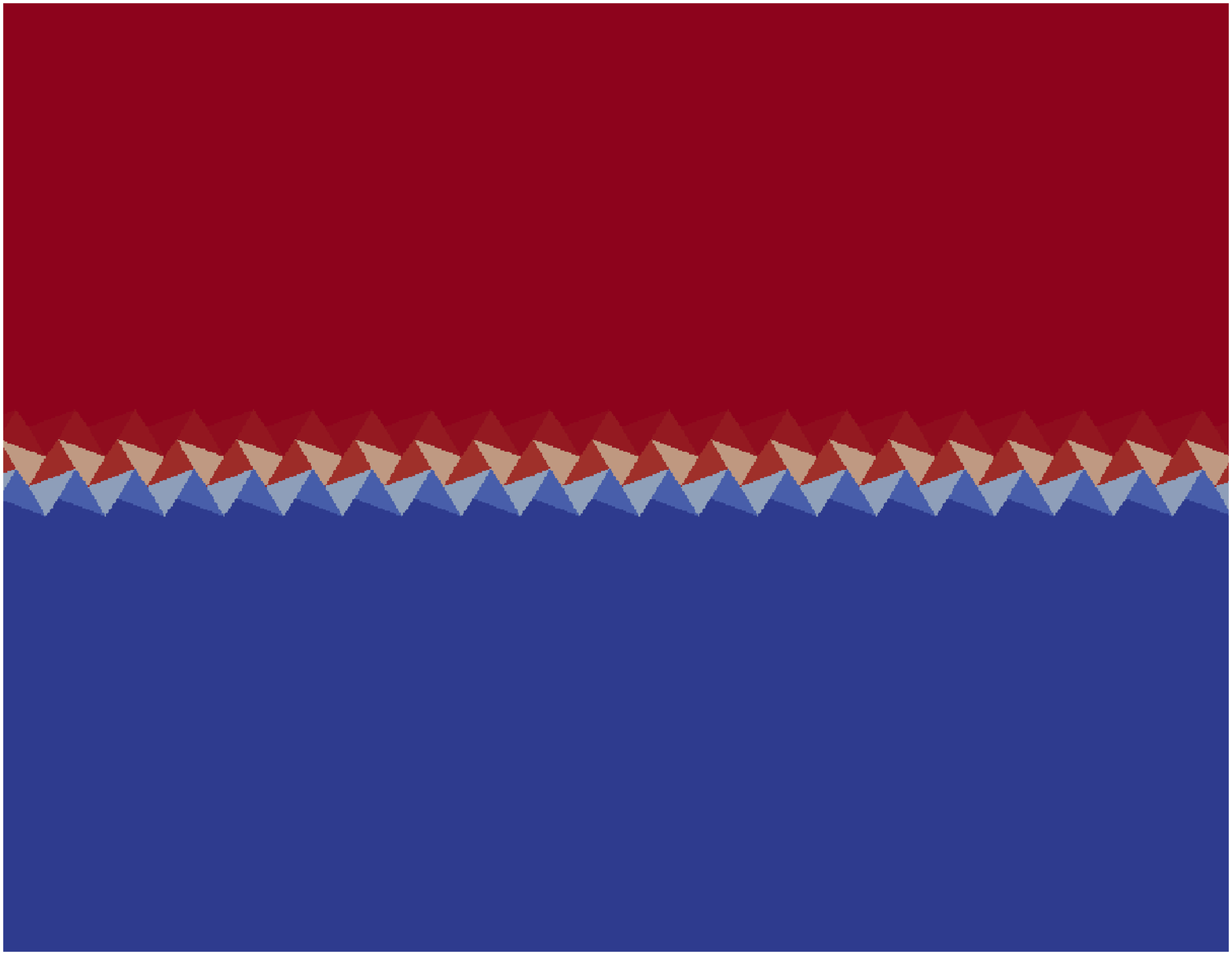}
	\end{subfigure}
	\begin{subfigure}[b]{0.19\textwidth}
		\centering
		\includegraphics[width=\textwidth]{figures/circleInitial.eps}
	\caption*{Initial state}
	\end{subfigure}
	\begin{subfigure}[b]{0.19\textwidth}
		\centering
		\includegraphics[width=\textwidth]{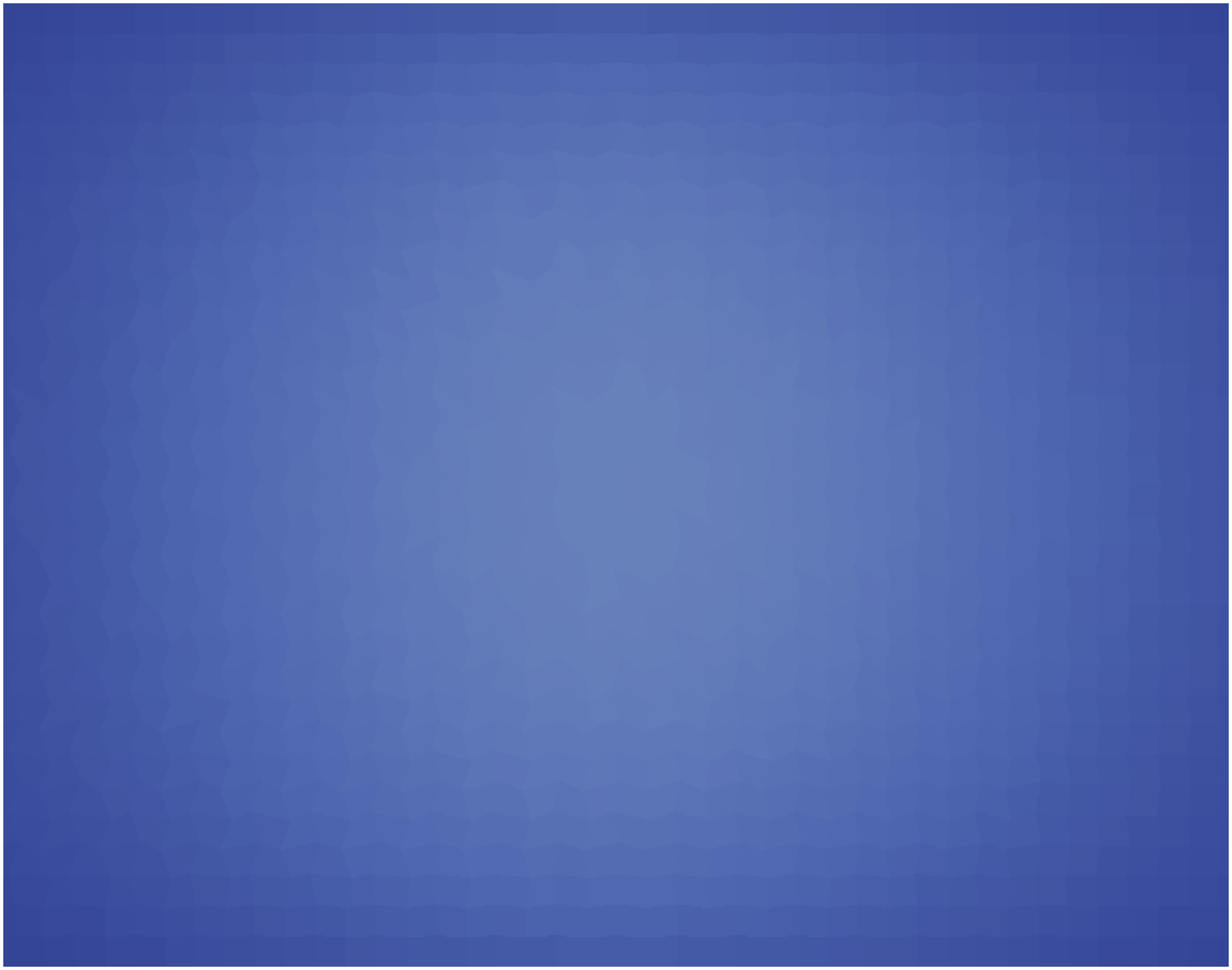}
	\caption*{UDS\,/\,SISC}
	\end{subfigure}
	\begin{subfigure}[b]{0.19\textwidth}
		\centering
		\includegraphics[width=\textwidth]{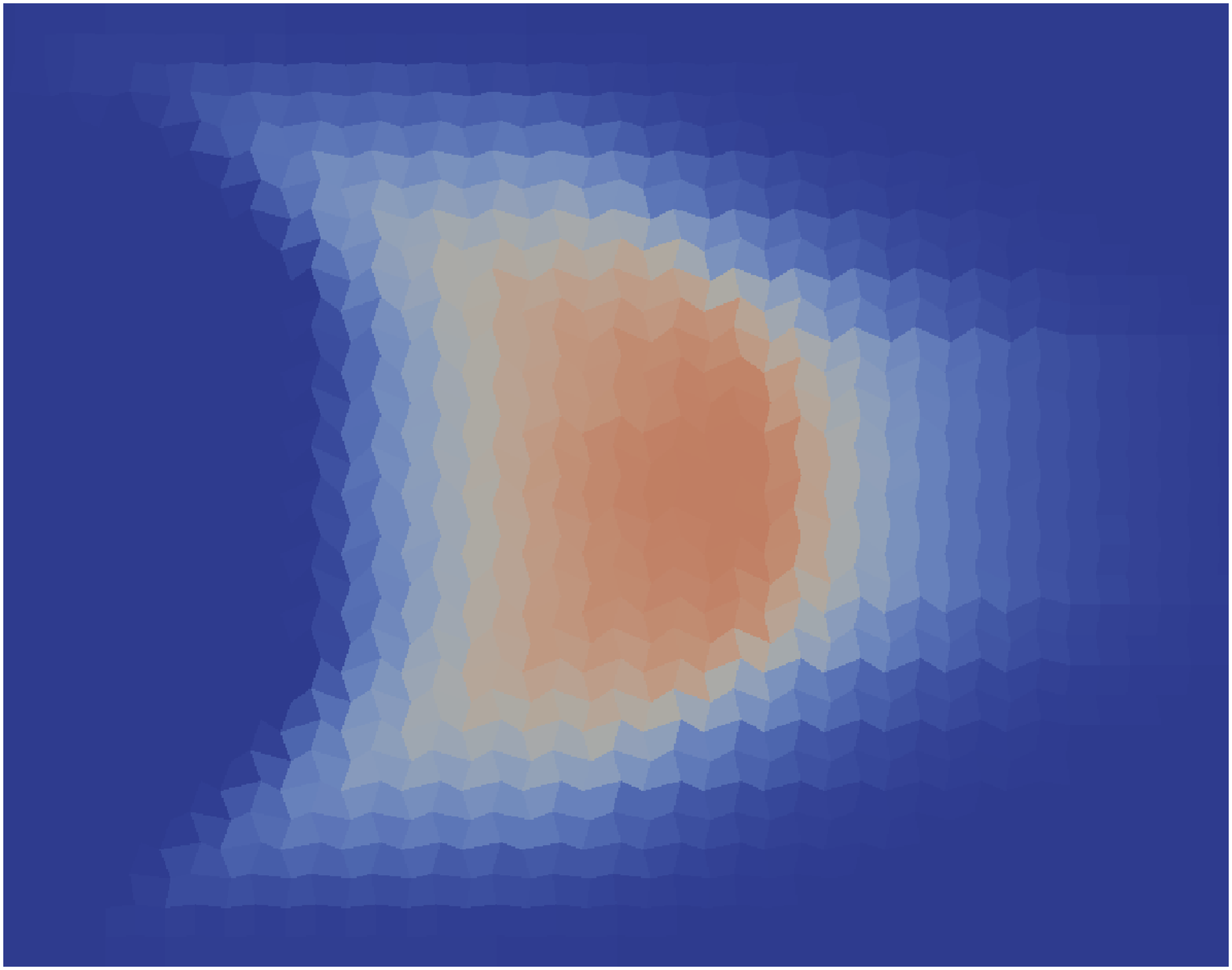}
	\caption*{CDS\,/\,SISC}
	\end{subfigure}
	\begin{subfigure}[b]{0.19\textwidth}
		\centering
		\includegraphics[width=\textwidth]{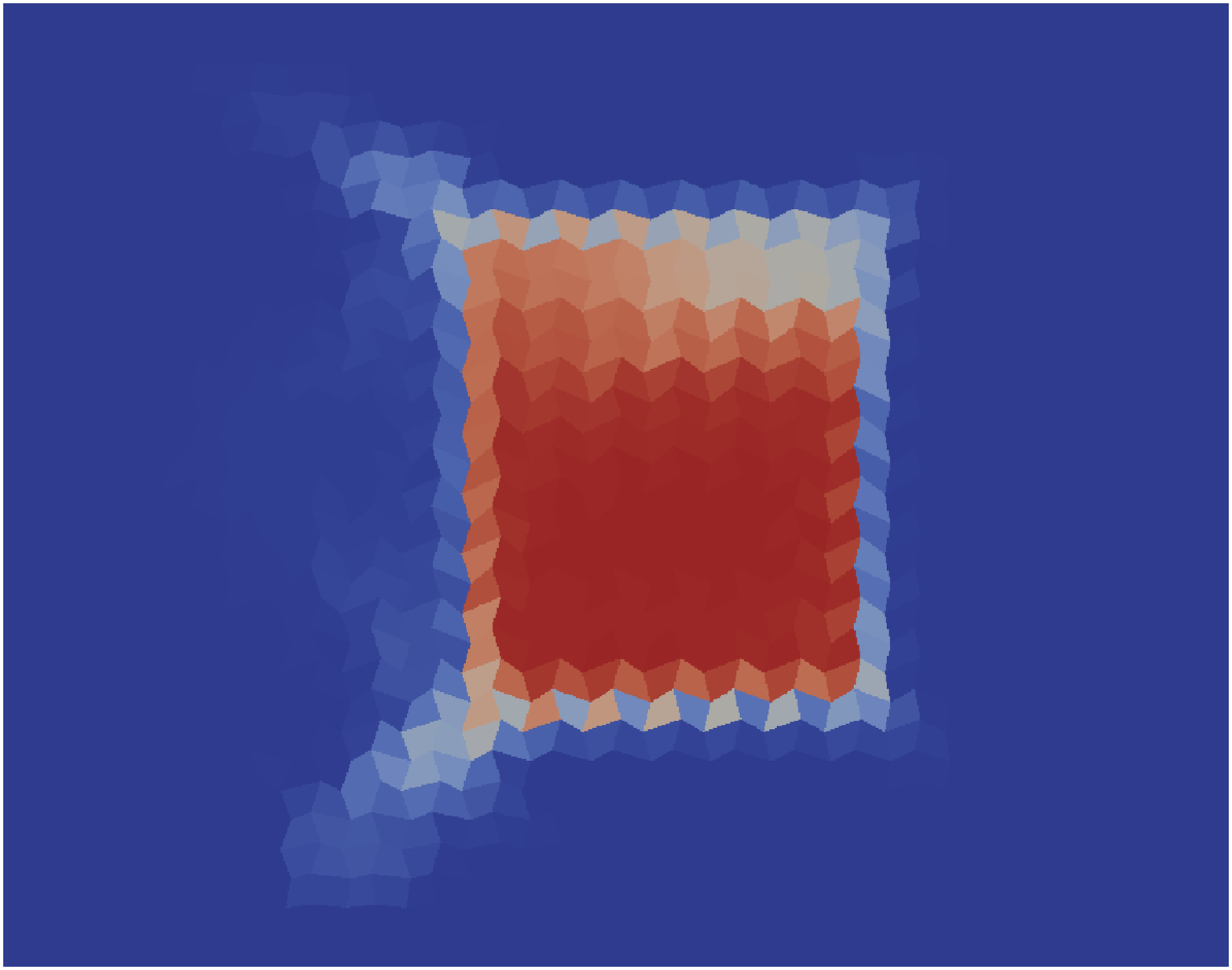}
	\caption*{CICSAM\,/\,SISC}
	\end{subfigure}
	\begin{subfigure}[b]{0.19\textwidth}
		\centering
		\includegraphics[width=\textwidth]{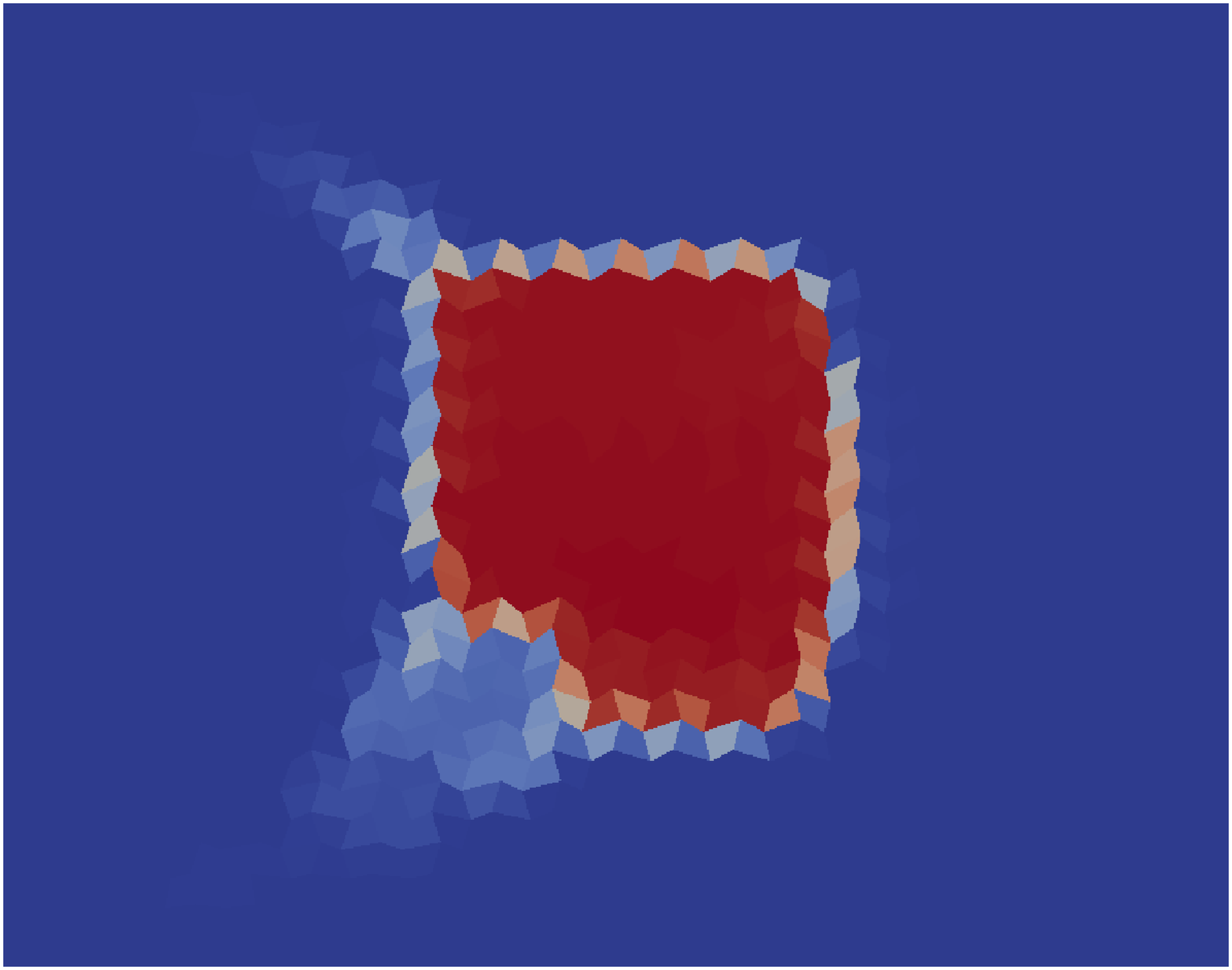}
	\caption*{LB\,/\,SISC}
	\end{subfigure}
    \caption{Visualisation of phase fraction field for the case of the applied \scheme{SISC} scheme in combination with different base schemes.}
	\label{figAdvectionSISCVisual}
\end{figure}                               

Looking at the distribution of the phase fraction obtained with the \scheme{SISC} scheme 
(depicted in Figure \ref{figAdvectionSISCVisual}) allows a more
extensive evaluation. The \scheme{UDS} scheme 
clearly shows a diffusive effect 
which is so pronounced that neither the shape of the plug flow nor the circular shape are preserved.
To a lesser degree, this effect is also visible when CDS is used. \scheme{CICSAM} and the \scheme{LB} 
scheme show good results in both
test cases. However, although \scheme{CICSAM} preserves sharp gradients quite well, the \scheme{LB} 
scheme is able to keep the initial 
steepness of the profile. Addressing the ability to preserve the circular shape it can be seen that both schemes tend 
to transform the circle into a quadratic shape, for the mesh topology under consideration.
Hence, for algebraic VoF methods the mesh topology has a major influence on accuracy (see Figure \ref{figAdvectionPolyHexa}).
Flux/cell-face alignment is important for the accuracy of algebraic VoF methods and thus polyhedral mesh topologies are to be favoured over to
quadrilateral (or hexahedral) meshes. Numerical simulations using standard discretisation on polyhedral meshes, however, 
suffer from mesh-induced skewness errors and hence the present work becomes generally important for algebraic VoF methods.
\begin{figure}[htb]
	\centering
	\begin{subfigure}[b]{0.25\textwidth}
		\centering
		\includegraphics[width=\textwidth]{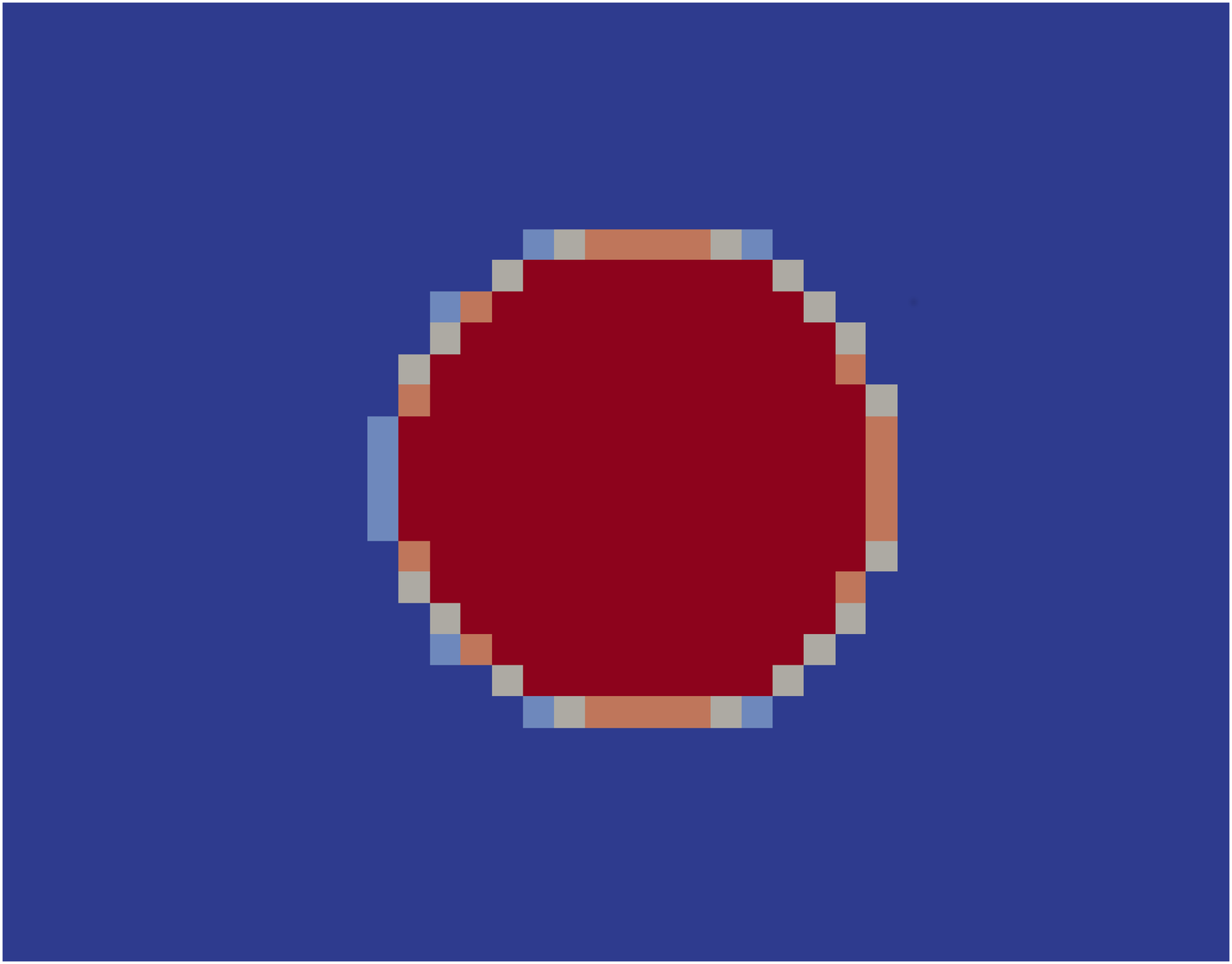}
	\end{subfigure}
	\begin{subfigure}[b]{0.25\textwidth}
		\centering
		\includegraphics[width=\textwidth]{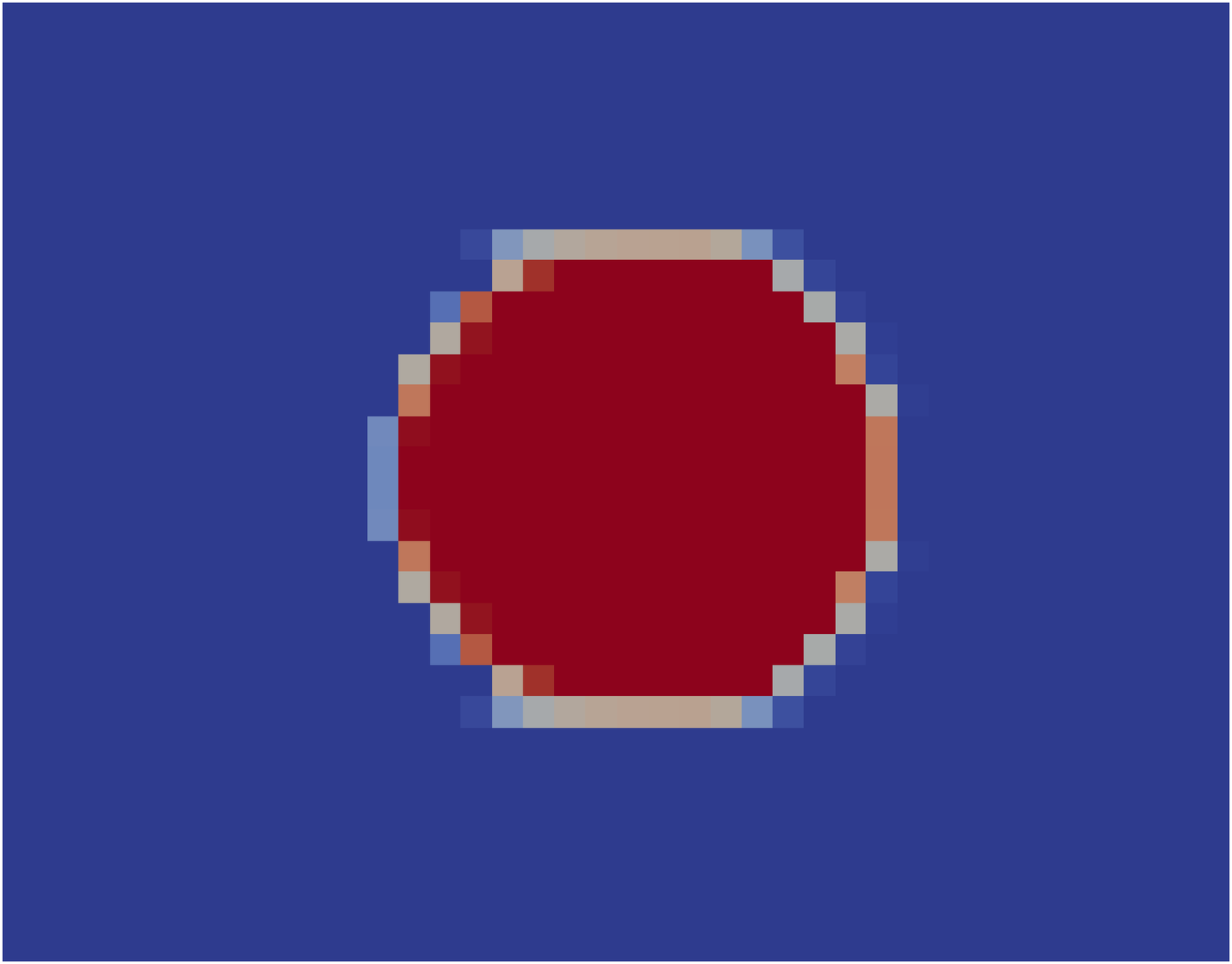}
	\end{subfigure}
	\begin{subfigure}[b]{0.25\textwidth}
		\centering
		\includegraphics[width=\textwidth]{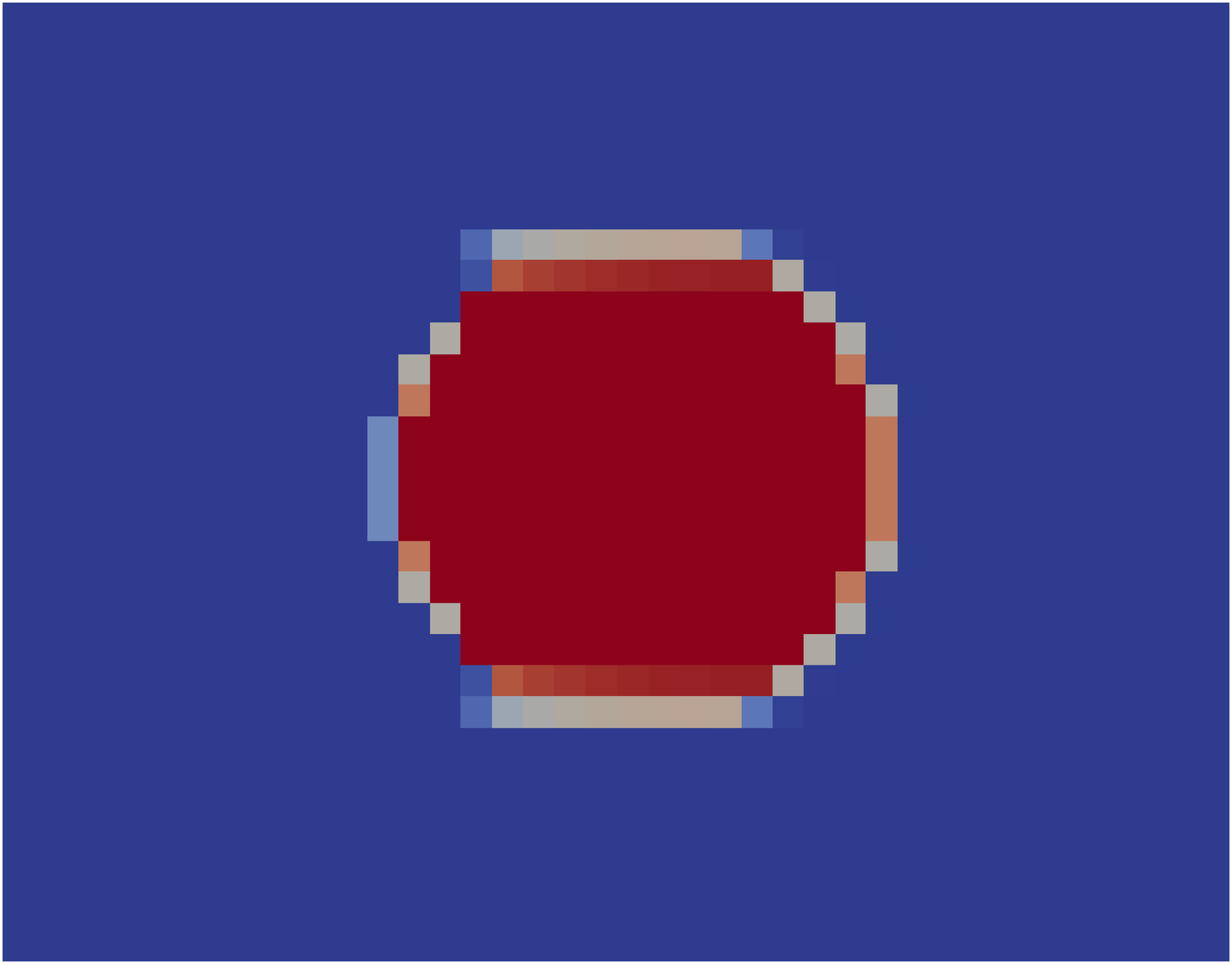}
	\end{subfigure}
	\begin{subfigure}[b]{0.25\textwidth}
		\centering
		\includegraphics[width=\textwidth]{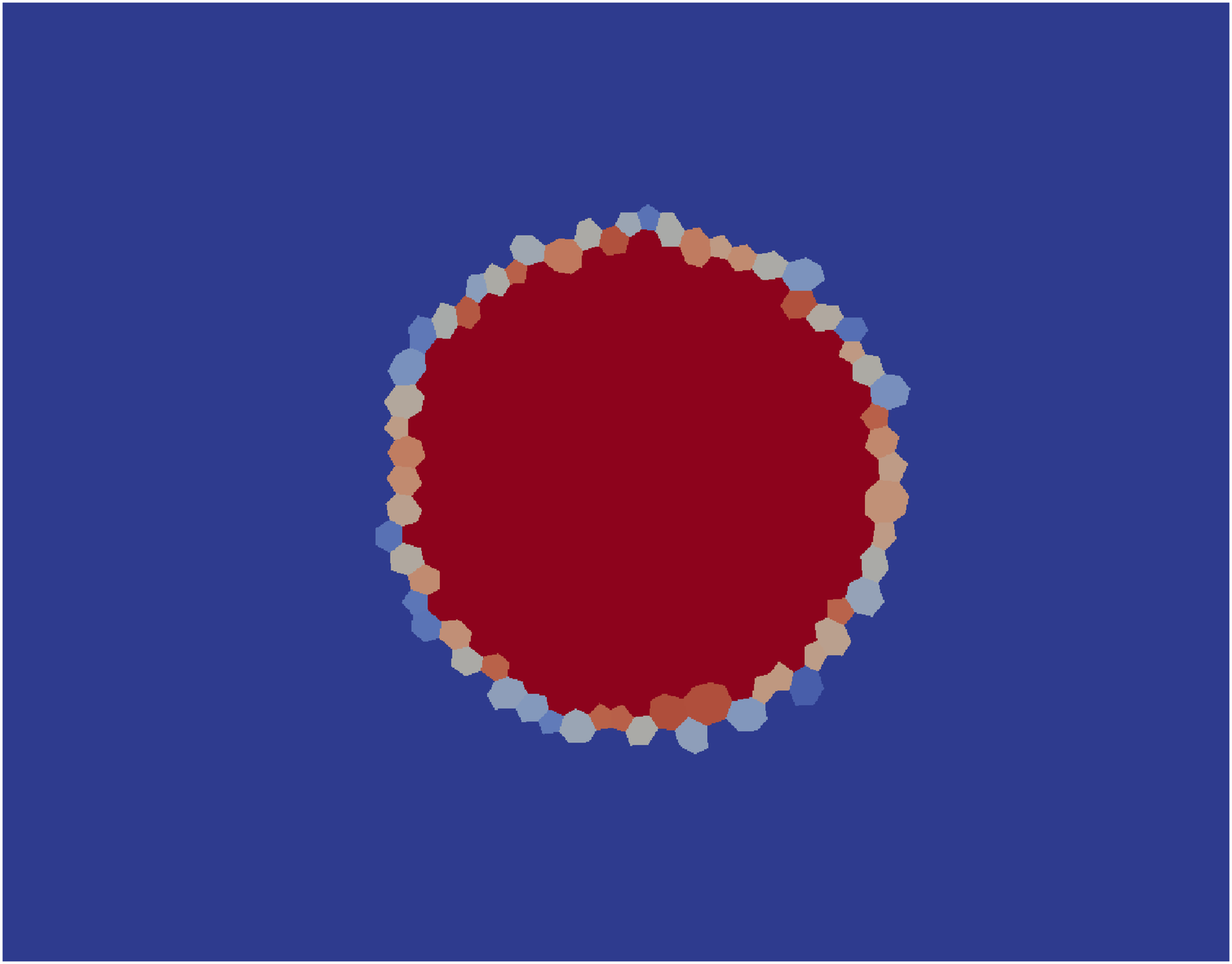}
	\caption*{Initial state}
	\end{subfigure}
	\begin{subfigure}[b]{0.25\textwidth}
		\centering
		\includegraphics[width=\textwidth]{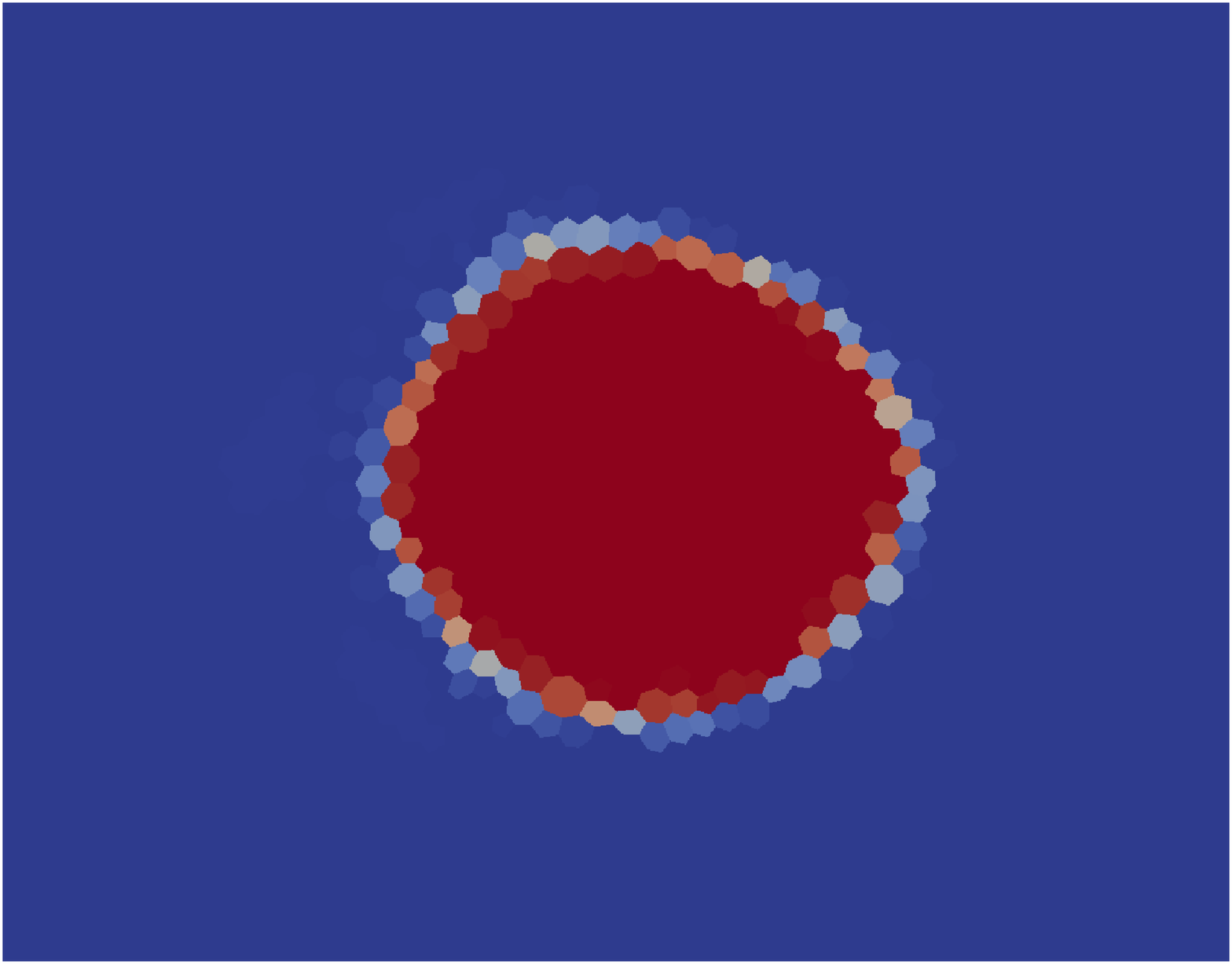}
	\caption*{CICSAM\,/\,UC}
	\end{subfigure}
	\begin{subfigure}[b]{0.25\textwidth}
		\centering
		\includegraphics[width=\textwidth]{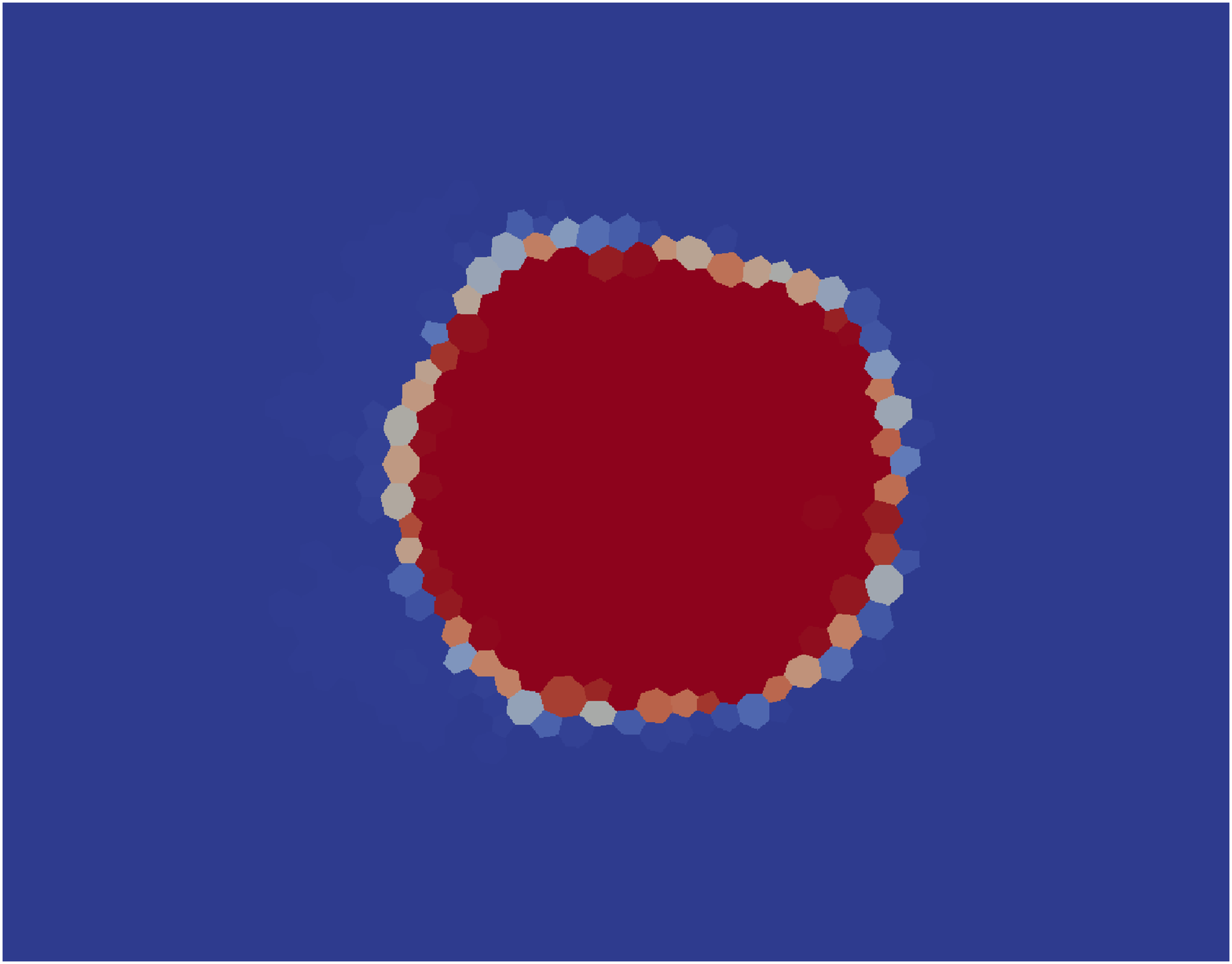}
	\caption*{LB\,/\,SISC}
	\end{subfigure}
    \caption{Resulting $\alpha$-distribution in case of applied \scheme{SISC} and \scheme{CICSAM} scheme when using a uniform cartesian
             (upper figures) and polyhedral (lower figures) mesh.}
	\label{figAdvectionPolyHexa}
\end{figure} 

To conclude it can be stated that the proposed correction scheme, in contrast with all other tested schemes, is able to 
keep the boundedness of the solution and at the same time, if combined with a suitable interpolation scheme, is capable of 
producing accurate results.

\subsection{Correction of CST-based species transfer term}
Different variants for discretising 
the terms of the CST-model are tested, with special attention devoted to mesh induced error correction
(cf. Table~\ref{tabDiscretisationCST}).
The three main questions are: 
\begin{itemize}
    \item How severe is the impact of skewed meshes on the accuracy of the CST-model and which of the proposed
          correction strategies is needed to restore the expected order of accuracy on non-distorted meshes?
    \item Does explicit correction lead to unboundedness of the solution? In other words, is implicit limiting necessary?
    \item Is it sufficient to account for non-orthogonality (NO) only, or must also non-conjunctional error correction (NO/NC) 
          be taken into account when discretising a gradient in face normal direction? 
\end{itemize}
Even though the last question is commonly answered in the negative it shall nevertheless be discussed here,
since species concentration fields commonly exhibit large discontinuities at the interface in both value and gradient. 
Consequently, errors introduced by skewed meshes are more pronounced and 
may necessitate a full skewness correction to obtain accurate results.

\subsubsection{Test Case}
In order to find the scheme which best meets the requirements, 
the simulation results of a planar diffusion case are evaluated.
To avoid effects of the convective term the velocity field is set to zero.
To evaluate the different correction strategies, species transfer from a surrounding stagnant gas 
into a stagnant liquid film is considered (see Figure \ref{figPlanarDiffusionSketch}). 
Five different Henry coefficients ($\operatorname{H}= 0.033, 0.2, 1, 5, 30$) are tested, where the 
diffusion coefficient in the gas phase is $D_g=1\cdot10^{-1}\, \text{m}^2/\text{s}$ and  
in the liquid phase $D_l=1\cdot10^{-5}\, \text{m}^2/\text{s}$. The initial gas concentration is $c=1$ and 
the liquid phase is initialized with $c=0$.

\begin{figure}[htb]
    \centering
    \includegraphics[width=.50\textwidth]{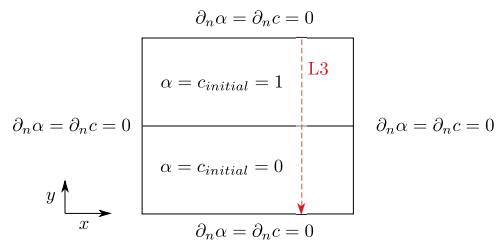}
    \caption{Set-up of test case to evaluate different corrected discretisations of the diffusive term. Here the species diffuses
             from the gas phase (upper half) into the liquid phase (lower half). The dashed red line indicates the
             path over which the concentration $c$ is plotted for evaluation reasons. The extension of the domain in the
             y-direction is $0.04$~m.}
    \label{figPlanarDiffusionSketch}
\end{figure}

In order to obtain the time-dependent numerical exact reference solution, the diffusion equation is discretised in 1D
using the Finite Difference Method (FDM). Here the same numerical set-up 
(domain size, boundary and initial conditions) is employed and the domain is sufficiently resolved. 
The solution which is obtained by implicitly solving the system of equations with Octave version 3.2.4 
is referred to later as Exact.

\subsubsection{Results}
Applying a specific correction when discretising the face-interpolated values 
(Table~\ref{tab:discrCSTb}), the results in some cases slightly improve, in others, however, deteriorate. 
A possible explanation is that to effectively correct the calculation of the face-interpolated 
values within a steep profile, a small gradient stencil and, at the same time, an accurate gradient computation are required. 
To maintain a small stencil, we utilize the Gaussian gradient for correction of the face-interpolation $\left(\text{K}c\right)_f$. 
However, as severely distorted meshes are used, the accuracy of the approximation suffers which is probably the cause for 
the observed poor performance. 
Consequently in the following only the variation of the gradient discretisation 
(Table~\ref{tab:discrCSTa}) is discussed while the face-interpolated values are discretised 
by simply using the uncorrected CDS. 

\begin{figure}[htb]
    \centering
    \graphicspath{{gnuplot/}}
    \resizebox{1.0\textwidth}{!}{\input{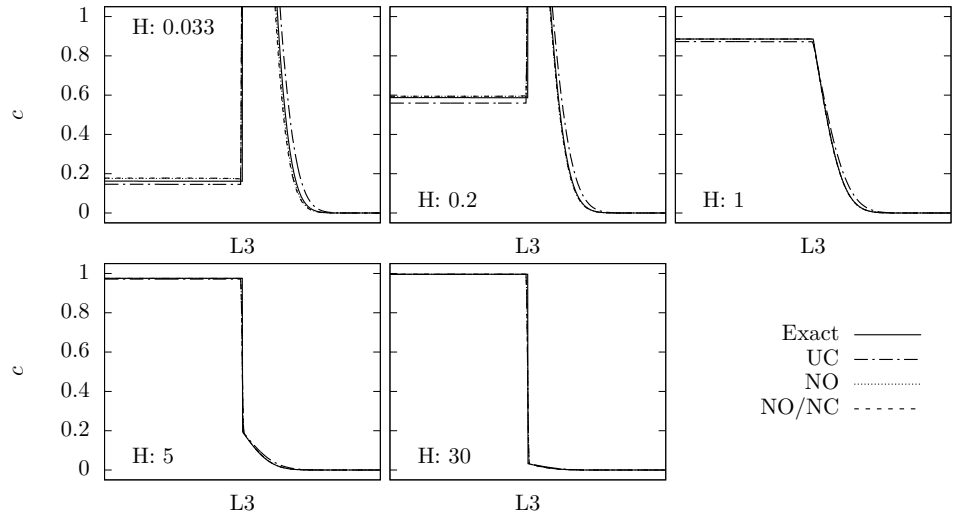}}
    \caption{Profile of concentration $c$ after time $t=0.5$~s for different Henry coefficients $\operatorname{H}$ 
    and correction strategies. The computational grid used is the systematically distorted 
    hexahedral mesh depicted in Figure \ref{figMesh} with an average cell width of $\approx 1.3\cdot10^{-4}$~m.}
    \label{figDiffusion1sec} 
\end{figure}

Results are visualized by plotting the profile of the concentration $c$ over the path `L3' (see Figure \ref{figPlanarDiffusionSketch}).
As shown in Figure \ref{figDiffusion1sec}, this is done for all considered Henry coefficients $\operatorname{H}$ 
and correction strategies. Despite the heavily distorted mesh, all approaches lead to reasonable results.
For all Henry coefficients,
the uncorrected approach tends to overestimate the diffusion and therefore does not correspond to the 
exact result. 
However, \scheme{NO} correction and \scheme{NO/NC} correction, result in a good approximation to the exact solution indicating that both approaches
effectively correct the mesh-skewness induced errors (see Figure~\ref{figConcProfileSpy}). 

\begin{figure}[htb]
    \centering
	\begin{subfigure}[b]{0.45\textwidth}
		\centering
		\includegraphics[width=\textwidth]{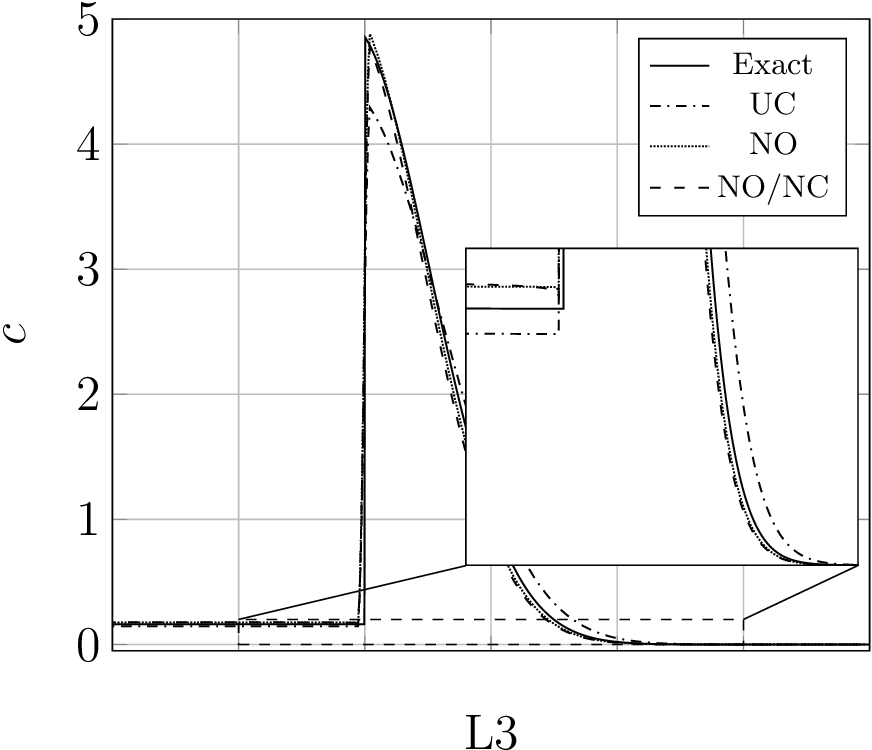}
	\caption{$\operatorname{H}$=$0.033$.}
	\end{subfigure}
	\begin{subfigure}[b]{0.45\textwidth}
		\centering
		\includegraphics[width=\textwidth]{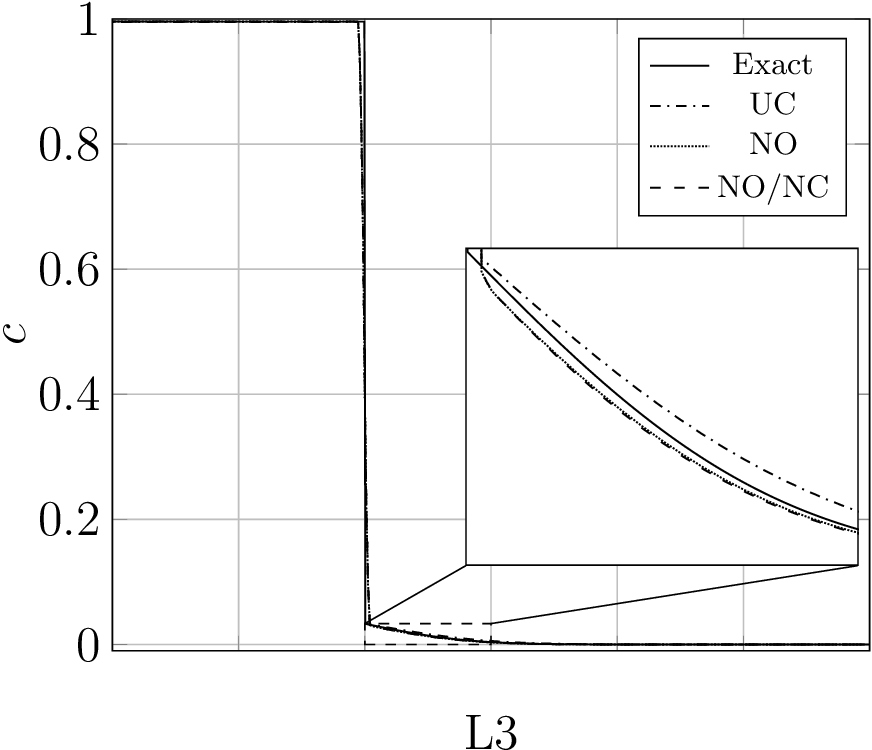}
	\caption{$\operatorname{H}$=$30$.}
    \label{spyHe30}
	\end{subfigure}
    \caption{Detailed view of concentration profiles for different Henry coefficients after time $t=0.5$~s. 
             The average cell width is $\approx 1.3\cdot10^{-4}$~m.}
    \label{figConcProfileSpy}
\end{figure}
Enlarged sections of two representative concentration profiles are shwon in Figure~\ref{figConcProfileSpy}.
It is shown that even for the lowest and highest Henry coefficients, no
unboundedness occured, which indicates that no limiting of the correction terms is required. 
However, one might intuitively try limiting the diffusive coefficient in the NO/NC formulation 
(cf. Equation~\ref{eqnSnGradNOSkewImplicit}), which is also tested in the present work. Due to the lack of a mathematically sound limiter criterion, 
the most intuitive one is applied, which is to restrict the modified diffusion coefficient to positive values. When applied, the results
deteriorated and the solution showed flickering in time. Based on this finding and the fact that the limiting 
does not seem to be crucial no further investigations were carried out and all the results presented below do not include any limiting. 

\begin{figure}[htb]
    \centering
    \graphicspath{{gnuplot/}}
    \resizebox{1.0\textwidth}{!}{\input{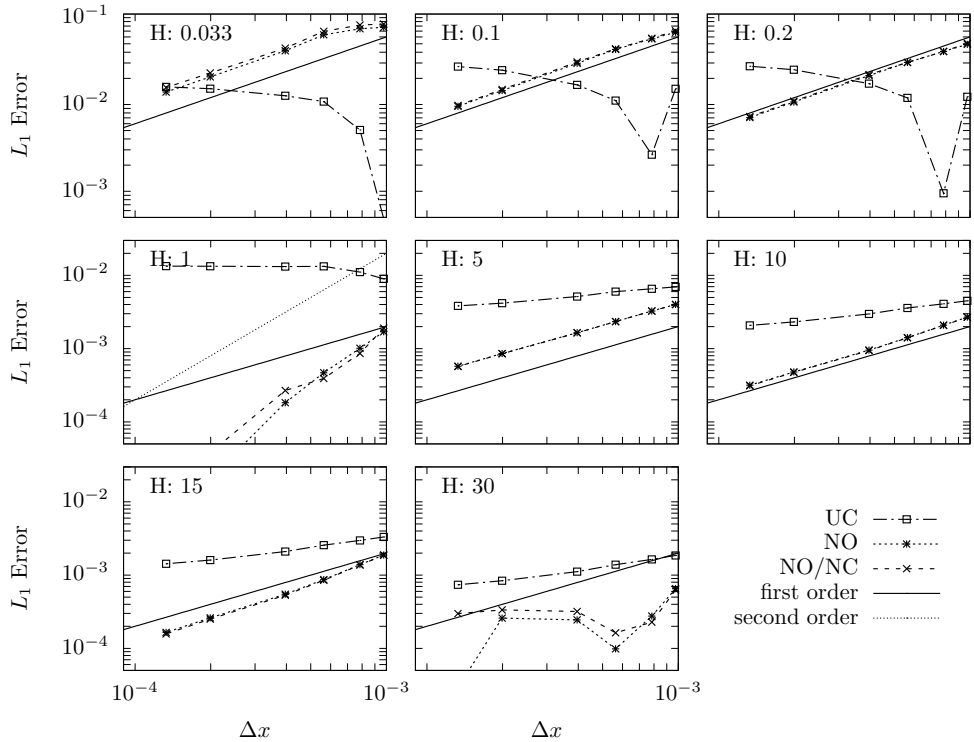}}
    \caption{Mesh convergence study at $t=0.5$~s}
    \label{figConvergenceStudyT05}                                                          
\end{figure}
To conclusively assess the convergence performance of the different schemes a mesh convergence study was conducted.
The average molar concentration in the gas phase after $t=0.1$~s and $t=0.5$~s was chosen as error measure . 
The latter is almost constant as the diffusive coefficient in the gas phase exceeds that in the liquid phase by orders
of magnitude. By this error definition the focus of the evaluation is not on the representation of the exact 
concentration profile (which we have already investigated, see Figure \ref{figDiffusion1sec} and \ref{figConcProfileSpy}) 
but on the correct prediction of the time integrated species flux over the interface. 
Figures~\ref{figConvergenceStudyT05} and~\ref{figConvergenceStudyT01} show the $L_1$ error plotted over the 
average cell size for each Henry coefficient at different times $t$.
It can be seen that the skewness-correction approaches employed generally lead to a substantial improvement of 
accuracy and convergence order. In all cases at least first order, for $\operatorname{H}=1$ (heat transfer) even second order convergence is achieved.
In contrast the uncorrected discretization converges with a much lower order or exhibits no convergence at all. 

\begin{figure}[htb]
    \centering
    \graphicspath{{gnuplot/}}
    \resizebox{1.0\textwidth}{!}{\input{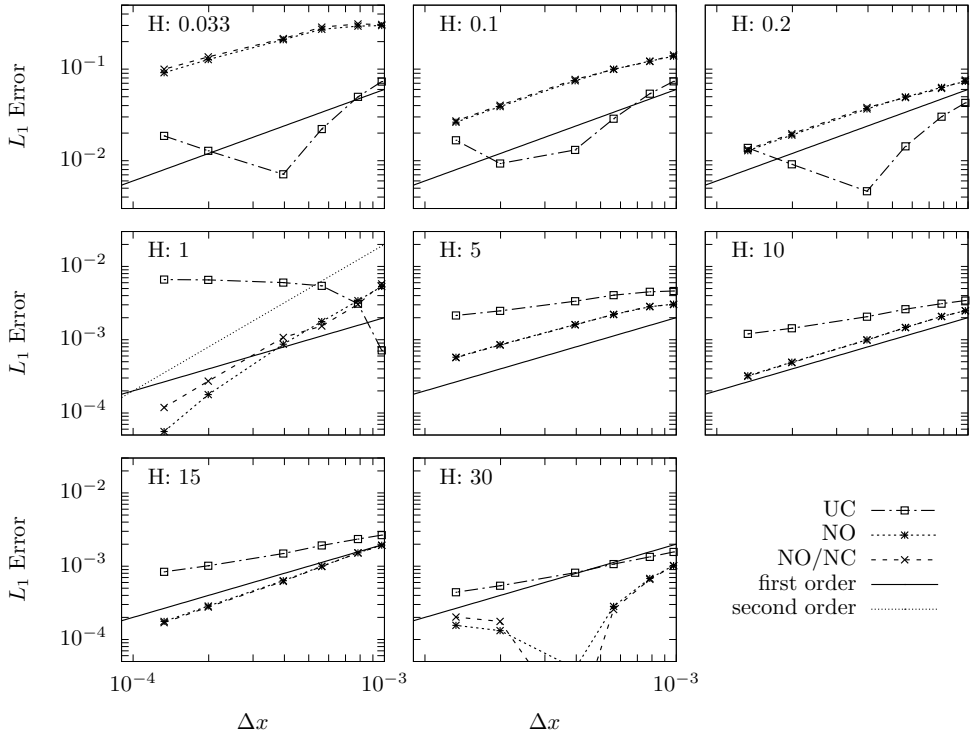}}
    \caption{Mesh convergence study at $t=0.1$~s}
    \label{figConvergenceStudyT01}                                                          
\end{figure} 
The observed convergence characterisic in the corrected case corresponds exactly to the formal order of the 
CST formulation, which is second order for $\operatorname{H}=1$ 
and tends to first order for $\operatorname{H}\ll 1$ and $\operatorname{H}\gg 1$
even on uniform cartesian meshes. It therefore can be stated that applying the corrections restores the \emph{formal} order of convergence 
on skewed meshes.
  
In the uncorrected case if $\operatorname{H}\leq 1$ and in the corrected case if $\operatorname{H}=30$,
unusual convergence behaviour is detected, that is the plot shows a local minimum of error. 
This can be explained by the fact that in these cases the concentration profile crosses the exact one when beeing evaluated using 
different mesh resolutions.
In consequence a minimum error is found at some random intermediate mesh resolution and seems to converge for lower and  
to diverge for higher resolutions. 
In addition, these cases obviously converge to a wrong solution. 
The magnitude of the absolute error thus corresponds to the $L_1$ error value it converges to for further refined meshes.

\begin{figure}[htb]
    \centering
	\begin{subfigure}[b]{0.30\textwidth}
		\centering
		\includegraphics[width=\textwidth]{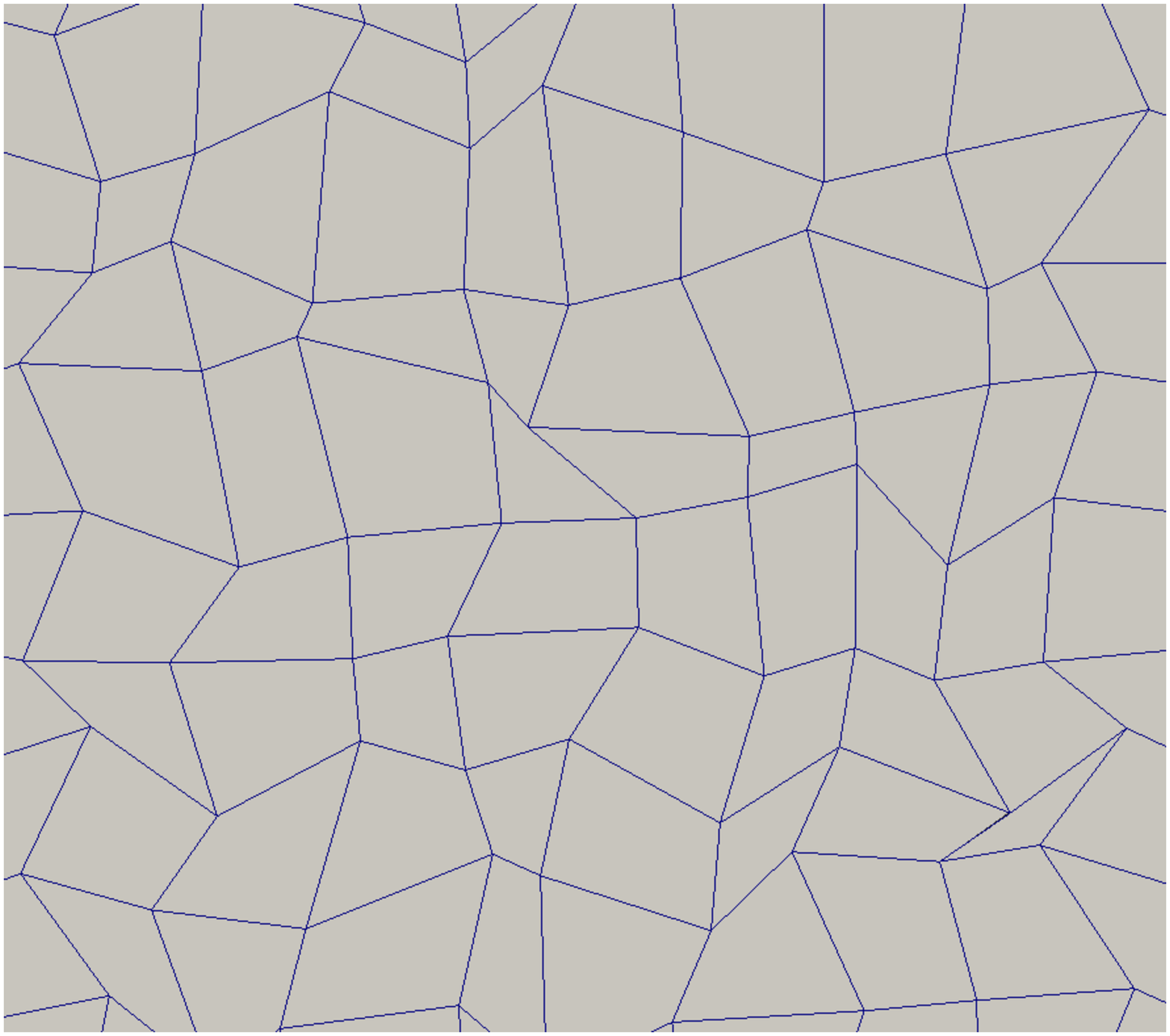}
	\caption{Distorted quadrilateral mesh.}
        \label{figMeshRandomHex}
	\end{subfigure}
	\begin{subfigure}[b]{0.30\textwidth}
		\centering
		\includegraphics[width=\textwidth]{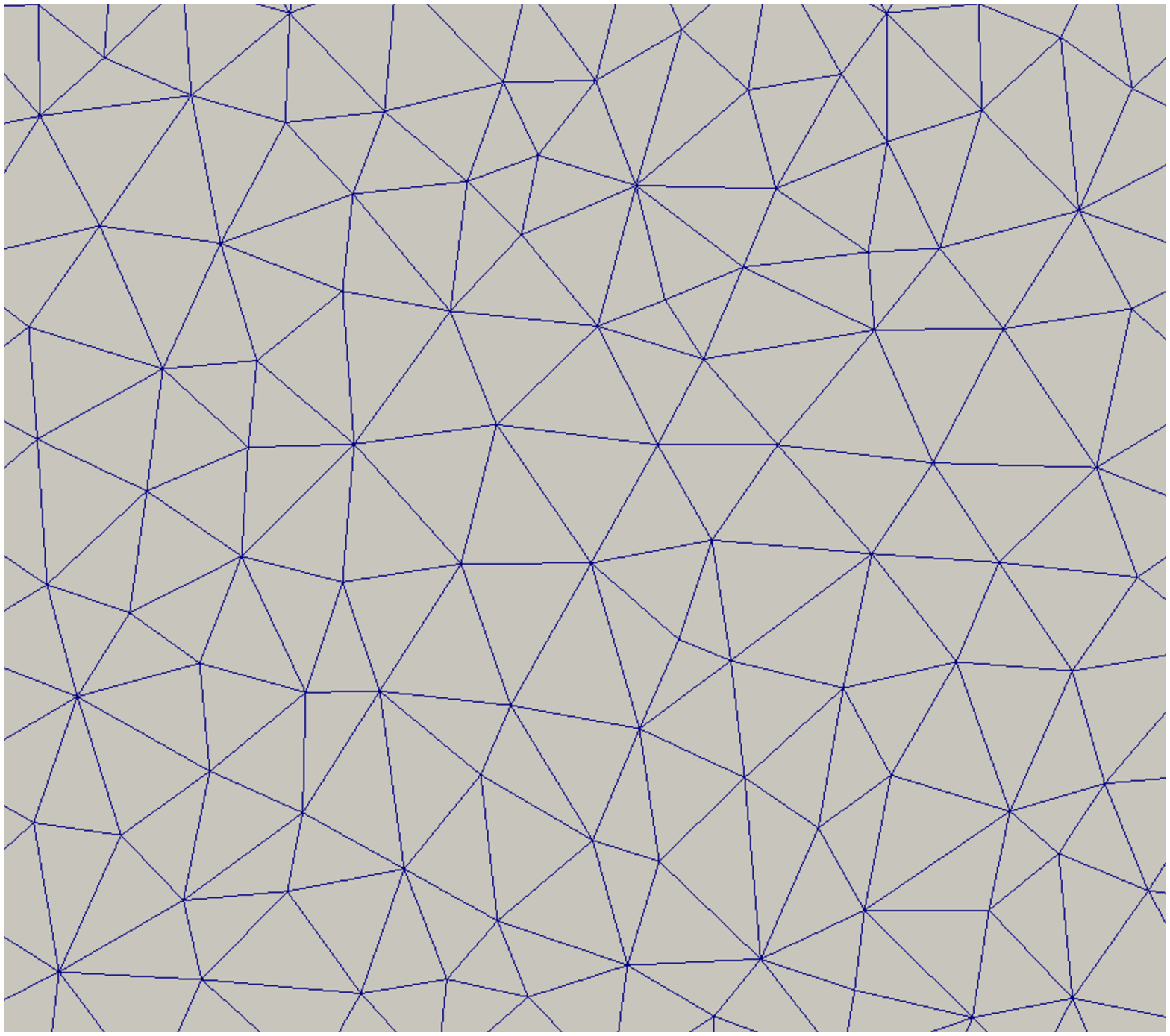}
	\caption{Randomly distorted tetrahedral mesh.}
        \label{figMeshRandomPoly}
	\end{subfigure}
    \caption{Structure of additionally generated meshes in order to further test correction of diffusion equation.}
    \label{figMeshAlternative}
\end{figure}
By considering both the convergence study and also the concentration profiles depicted in Figure~\ref{figDiffusion1sec}, 
the question about the impact of non-conjunctionality errors when discretising
diffusive terms can finally be answered: it is indeed negligible, since neither the NO nor the NO/NC profiles
differ significantly, even for cases where sharp gradients are obviously present. 

\begin{figure}[htb]
    \centering
    \graphicspath{{gnuplot/}}
    \resizebox{1.0\textwidth}{!}{\input{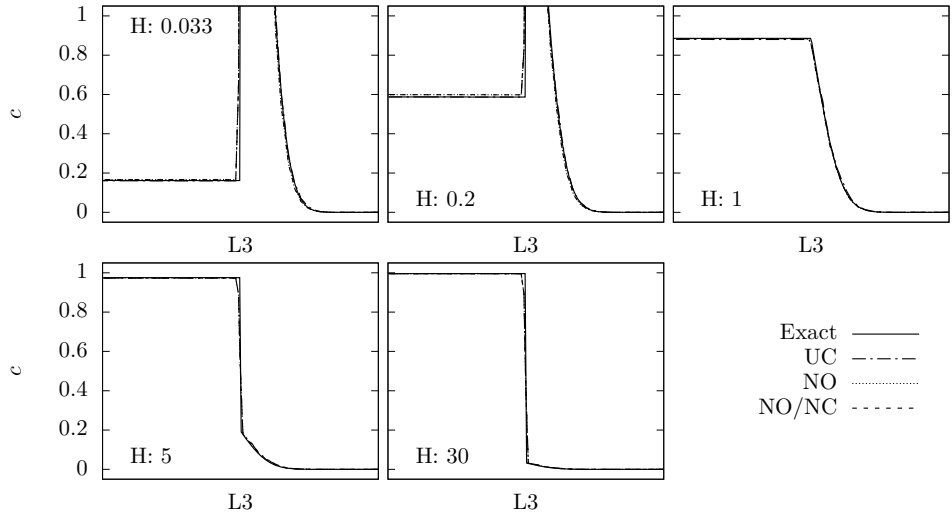}}
    \caption{Profile of concentration $c$ after time $t=0.5$~s for different Henry coefficients $\operatorname{H}$ 
    and correction strategies. The computational grid used is the randomly distorted hexahedral mesh 
    depicted in Figure \ref{figMeshRandomHex} with an average cell width of $\approx 4\cdot10^{-4}$~m.}
    \label{figDiffusionRandom1sec} 
\end{figure}
\begin{figure}[htb]
    \centering
    \graphicspath{{gnuplot/}}
    \resizebox{1.0\textwidth}{!}{\input{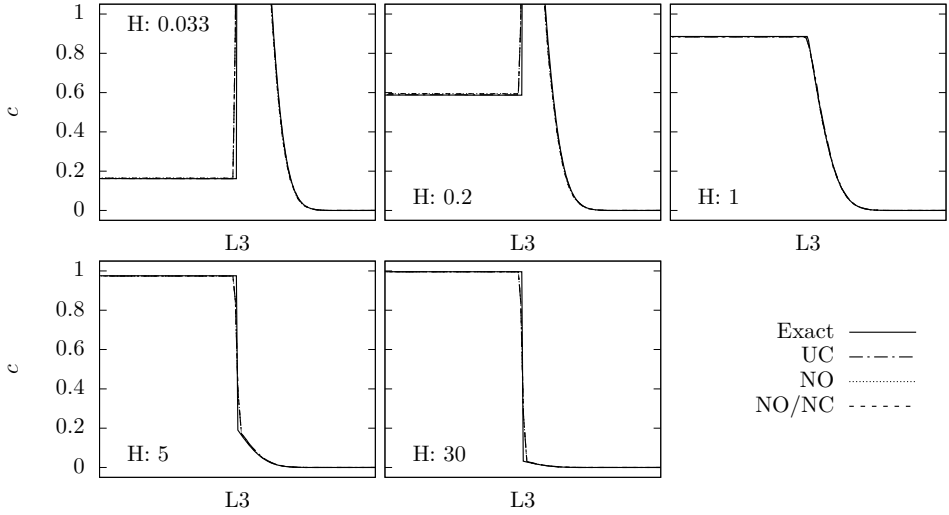}}
    \caption{Profile of concentration $c$ after time $t=0.5$~s for different Henry coefficients $\operatorname{H}$ 
    and correction strategies. The computational grid used is the tetrahedral mesh depicted in Figure \ref{figMeshRandomPoly}.
    with an average cell width of $\approx 4\cdot10^{-4}$~m.}
    \label{figDiffusionTetra1sec} 
\end{figure}
Considering the heavily distorted mesh, the achieved solution accuracy might be surprising. Hence, we also raise the question as to whether 
errors may possibly be cancelled out because of the use of a systematically distorted mesh (see Figure \ref{figMesh}). To eliminate this possibility 
all test cases are additionally solved on a randomly distorted hexahedral and tetrahedral mesh, both depicted in 
Figure \ref{figMeshAlternative}, as these are often used in literature.   

Figures \ref{figDiffusionRandom1sec} and \ref{figDiffusionTetra1sec} show the results analogous to the already discussed set of
graphs in Figure \ref{figDiffusion1sec}. In both cases, independent of the selected discretisation method or Henry coefficient, 
the results almost perfectly match the solution, which leads to the conclusion that randomly distorted meshes
(unlike the systematically distorted meshes studied here) result in error cancelation and are thus inappropriate for the systematic 
investigation of skewness correction approaches.

\section{Summary and Conclusions}
\label{sec:conclusions}
\vspace{-2pt}
This work is aimed at the targeted investigation and enhancement of mesh-skewness correction approaches with special focus on implicit
and bounded corrections in the context of VoF-based interfacial heat and mass transfer. 
The presence of a jump and/or steep gradients substantiate a sound reconsideration of the skewness correction approaches, 
which are commonly recognised as accepted knowledge in the literature. Discretisation procedures for the advection and diffusion terms 
are verified separately using different test cases on systematically distorted meshes. Randomly distorted meshes are shown to 
lead to error cancelation and thus are proven to be inappropriate for a systematic study of mesh-induced errors.

For the phase-fraction advection term we investigate the boundedness and accuracy achieved with different discretisation 
strategies, i.e.~the shape-preservation and sharpness properties of different approaches to skewness correction. The main 
finding is that implicit correction using the novel semi-implicit skewness correction (\scheme{SISC}) approach is imperative. 
Further, it is shown that the gradient scheme used for computation of the correction term has a strong influence on accuracy. 
In our study, the performance of the least square gradient computation is found superior over Gauss gradient approximation.

For the diffusive heat/species transfer terms in the CST method, again we investigate the boundedness and accuracy achieved 
with different discretisation strategies. Additionally, focus is on convergence on strongly distorted meshes.
It is shown that the approach for semi-implicit skewness correction (\scheme{SISC}) used for the advection term
can be adopted for diffusion terms as well. However, for the test case studied here applying a relatively small 
time step, boundedness of the diffusion term is also attained using explicit correction approaches, although not 
a-priori guaranteed as in the implicit approach. Non-orthogonality correction (\scheme{NO}) has been found to perform 
reasonably well even for the transport of fields with discontinuities. Simultaneuos implicit correction of 
non-orthogonality and non-conjunctionality errors (\scheme{NO}/\scheme{NC}) has shown no further benefit even for quantities 
exhibiting strong solution curvature, i.e.~strong spatial changes in the gradient.
Finally, the authors have demonstrated that the convergence order on non-distorted meshes can be retained even 
on strongly distorted meshes only when suitable correction approaches such as \scheme{NO} or \scheme{NO}/\scheme{NC} are used.

To conclude, it is shown that the CST-method is able to predict accurate solutions on moderatly distorted meshes even without
corrections. For severe distortions the accuracy can still be maintained by applying suitable, i.e.\ bounded implicit, mesh-skewness corrections.

\bibliographystyle{unsrt}
\bibliography{ms}
\end{document}